\documentclass[sigconf]{acmart}

\usepackage{booktabs} % For formal tables
\usepackage{subfigure}
\usepackage{color}

\setcopyright{acmcopyright}
\begin{document}

\copyrightyear{2019} 
\acmYear{2019} 
\acmConference[CIKM '19]{The 28th ACM International Conference on Information and Knowledge Management}{November 3--7, 2019}{Beijing, China}
\acmBooktitle{The 28th ACM International Conference on Information and Knowledge Management (CIKM '19), November 3--7, 2019, Beijing, China}
\acmPrice{15.00}
\acmDOI{10.1145/3357384.3358030}
\acmISBN{978-1-4503-6976-3/19/11}

%\settopmatter{printacmref=true}

\title{A Hierarchical Self-Attentive Model for Recommending User-Generated Item Lists}

\author{Yun He, Jianling Wang, Wei Niu and James Caverlee}
%\authornote{Dr.~Trovato insisted his name be first.}
%\orcid{1234-5678-9012}
\affiliation{%
  \institution{Department of Computer Science and Engineering, Texas A\&M University}
  %\streetaddress{P.O. Box 1212}
  %\city{Dublin}
  %\state{Ohio}
  %\postcode{43017-6221}
}
\email{{yunhe, jlwang, caverlee}@tamu.edu; weiniu.2010@gmail.com}

%
%\author{G.K.M. Tobin}
%\authornote{The secretary disavows any knowledge of this author's actions.}
%\affiliation{%
%  \institution{Institute for Clarity in Documentation}
%  \streetaddress{P.O. Box 1212}
%  \city{Dublin}
%  \state{Ohio}
%  \postcode{43017-6221}
%}
%\email{webmaster@marysville-ohio.com}
%
%\author{Lars Th{\o}rv{\"a}ld}
%\authornote{This author is the
%  one who did all the really hard work.}
%\affiliation{%
%  \institution{The Th{\o}rv{\"a}ld Group}
%  \streetaddress{1 Th{\o}rv{\"a}ld Circle}
%  \city{Hekla}
%  \country{Iceland}}
%\email{larst@affiliation.org}
%
%\author{Valerie B\'eranger}
%\affiliation{%
%  \institution{Inria Paris-Rocquencourt}
%  \city{Rocquencourt}
%  \country{France}
%}
%\author{Aparna Patel}
%\affiliation{%
% \institution{Rajiv Gandhi University}
% \streetaddress{Rono-Hills}
% \city{Doimukh}
% \state{Arunachal Pradesh}
% \country{India}}
%\author{Huifen Chan}
%\affiliation{%
%  \institution{Tsinghua University}
%  \streetaddress{30 Shuangqing Rd}
%  \city{Haidian Qu}
%  \state{Beijing Shi}
%  \country{China}
%}
%
%\author{Charles Palmer}
%\affiliation{%
%  \institution{Palmer Research Laboratories}
%  \streetaddress{8600 Datapoint Drive}
%  \city{San Antonio}
%  \state{Texas}
%  \postcode{78229}}
%\email{cpalmer@prl.com}
%
%\author{John Smith}
%\affiliation{\institution{The Th{\o}rv{\"a}ld Group}}
%\email{jsmith@affiliation.org}
%
%\author{Julius P.~Kumquat}
%\affiliation{\institution{The Kumquat Consortium}}
%\email{jpkumquat@consortium.net}
%
%% The default list of authors is too long for headers.
%\renewcommand{\shortauthors}{B. Trovato et al.}

\begin{abstract}
User-generated item lists are a popular feature of many different platforms. Examples include lists of books on Goodreads, playlists on Spotify and YouTube, collections of images on Pinterest, and lists of answers on question-answer sites like Zhihu. Recommending item lists is critical for increasing user engagement and connecting users to new items, but many approaches are designed for the item-based recommendation, without careful consideration of the complex relationships between items and lists. Hence, in this paper, we propose a novel user-generated list recommendation model called AttList. Two unique features of AttList are careful modeling of (i) hierarchical user preference, which aggregates items to characterize the list that they belong to, and then aggregates these lists to estimate the user preference, naturally fitting into the hierarchical structure of item lists; and (ii) item and list consistency, through a novel self-attentive aggregation layer designed for capturing the consistency of neighboring items and lists to better model user preference. Through experiments over three real-world datasets reflecting different kinds of user-generated item lists, we find that AttList results in significant improvements in NDCG, Precision@k, and Recall@k versus a suite of state-of-the-art baselines.  Furthermore, all code and data are available at https://github.com/heyunh2015/AttList.

\end{abstract}




\maketitle

\section{Introduction}
\label{Introduction}
User-generated item lists are a widespread feature of many platforms. Examples include music playlists on Spotify, collections of videos on YouTube, lists of books on Goodreads, wishlists of products on Amazon, collections (boards) of images (pins) on Pinterest, and lists of interesting answers on question-answer sites like Zhihu. These item lists directly power user engagement -- for example, more than 50\% of Spotify users listen to playlists, accounting for more than 1 billion plays per week \cite{buzzfeed2016}; and Pinterest users have curated more than 3 billion pins to about 4 billion boards \cite{eksombatchai2018pixie}.

\begin{figure}[]
    \centering
    \setlength{\abovecaptionskip}{0.01cm}
    \setlength{\belowcaptionskip}{-0.2cm}
    \includegraphics[scale=0.25]{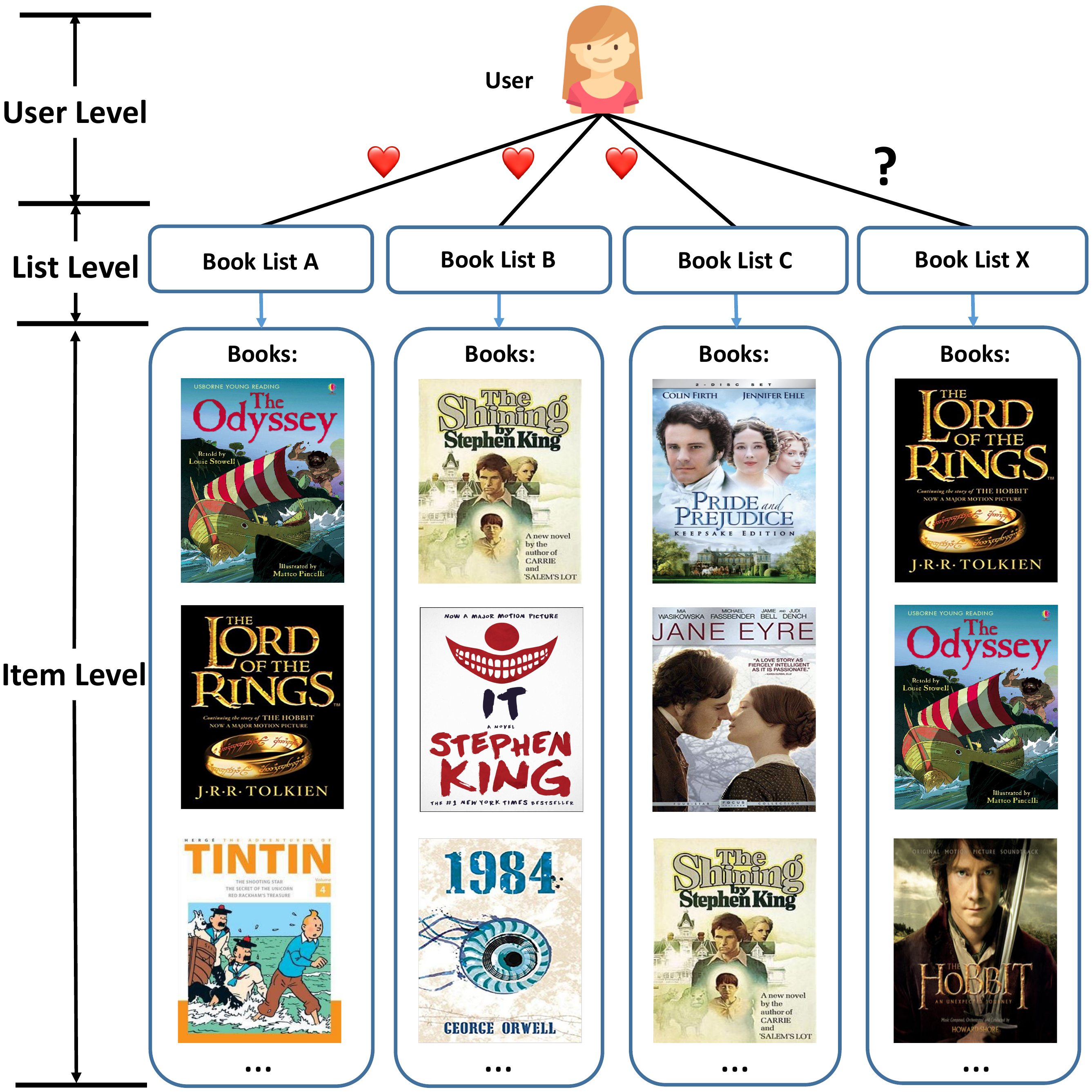}
    \caption{Example: A Goodreads user likes three different lists (A, B, and C), each composed of a collection of books. Our proposed model exploits this user-list-item hierarchical structure to recommend additional lists (e.g., X).
   	}
    \label{toy data}
\end{figure}

%curate pins to one of their boards 

%a typical example is that users on Pinterest must curate a pin into one of their boards (more than 4 billion \cite{eksombatchai2018pixie}) if they want to save the pin. Another example is that more than 50\% of Spotify users listen to playlists, accounting for more than 1 billion plays per week \cite{buzzfeed2016}. Besides, users on Netease create 0.6 million new playlists daily and there are more than 400 million playlists totally. \cite{Netease2017}

%For example, playlists accounted for 31\% of music listening time among listeners in the United States, which is more than albums (22\%)\footnote{https://musicbiz.org/news/playlists-overtake-albums-listenership-says-loop-study/}

Across platforms, user-generated item lists are manually created, curated, and managed by users, providing a unique perspective into how items (e.g., songs, videos, books, and answers) can be grouped together. Lists may be organized around theme, genre, mood, or other non-obvious pattern, so users can easily explore and consume correlated items together. To illustrate, Figure \ref{toy data} presents an example of several lists on Goodreads, a book sharing platform. This user likes book lists organized around different themes (e.g., adventure and horror). 

But how can we effectively connect users to the right lists? Traditional \textit{item-based recommendation} has shown good success in modeling user preferences for specific items \cite{koren2009matrix, hu2008collaborative}. In contrast, \textit{list-based recommendation} is a nascent area with key challenges in modeling the complex interactions between users and lists, between lists and items, and between users and items. For example, a user's preference for a list may be impacted by factors such as the overall theme of the list, all of the items or just a few of the items on the list, the position of items on the list, and so on. The user in Figure \ref{toy data} may prefer list A for its classic mix of young adult adventure stories, but only like list B for its first two horror novels (by the same author Stephen King), regardless of the rest of the items on the list. Furthermore, the variety across book lists liked by this user raises tough challenges in terms of modeling preference for new lists (as in the unknown list X). A few previous efforts \cite{liu2014recommending,cao2017embedding} have proposed to treat user-list and user-item interactions through a collective matrix factorization approach \cite{singh2008relational}, where a user-list interaction matrix and a user-item matrix are jointly factorized  \cite{liu2014recommending,cao2017embedding}. However, it is non-trivial to weight user-list and user-item interactions in the loss function for optimizing the list recommendation task. The number of user-item interactions are often much more than user-list interactions and thus dominate the loss. This imbalance in the number of user-list and user-item interactions stems from the hierarchical structure of item lists, i.e., a list often contains tens or even hundreds of items, as shown in Section \ref{Data Sets}.

Hence, with these challenges in mind, we propose a novel hierarchical self-attentive model for recommending user-generated item lists that is motivated by two key observations:

%Through careful observations, we find out that a hierarchical aggregated representation framework can be naturally fitted into the hierarchical structure of user-generated item lists.

\medskip
\noindent\textbf{Hierarchical user preferences.} First, users and user-generated item lists naturally form a hierarchical structure that could be helpful for modeling user preferences. To illustrate, Figure \ref{toy data} shows how a user (from the top) can like a collection of different lists, which in turn are composed of different items (at the bottom). Hence, it is natural to represent lists by their constituent items, and represent users by the lists that they prefer. Our first motivating observation then is \textit{user preferences can be propagated bottom-up from the item level to the list level and finally to the user level.}

\medskip
\noindent\textbf{Item and list consistency.} Second, the similarity between items and their neighboring items (whether by genre, theme, or other pattern) is a key clue for modeling the importance of those items towards reflecting user preferences. Likewise, the similarity between a list and its neighboring lists can reveal the importance of those lists to the user's overall preferences. We refer to this similarity as \textit{item consistency} and \textit{list consistency}. For example, if a scary book (e.g., \textit{The Shining}) is curated in a list (like list C in Figure \ref{toy data}) mainly composed of romance books (e.g., \textit{Pride and Prejudice}, \textit{Jane Eyre}), then it is less informative in terms of the overall list characteristics. Likewise, a romance list may be less informative about the user preference if it is not consistent with the rest of the user's profile. Our second motivating observation is \textit{the lower the consistency between an item and the rest of the items on a list, the less likely it can reveal the list's characteristics. And conversely, the higher the consistency, the more likely it can reveal the list's characteristics.} Similarly, the higher the consistency, the more likely a list can reveal the user's preference. That is, these inconsistent items and lists should be assigned lower weights to represent the list and the user.

%, and the similarity between a list and its neighboring lists 

% In this paper, consistency means the similarities between an item and its neighboring items in terms of genre, theme or other non-obvious pattern. 
 
 %Similarly, the lower the consistency between a list and other user's liked lists, the less is revealed about the user's hidden preference.

These observations motivate us to attack the user-generated item list recommendation problem with a \textit{hierarchical} user-list-item model that naturally incorporates a \textit{self-attentive} aggregation layer to capture item and list consistency by correlating them to their neighboring items and lists. In summary:

%Our contributions are further summarized as follows:
\begin{itemize}
	\item We study the important yet challenging problem of recommending user-generated item lists -- a complex recommendation scenario that poses challenges beyond existing item-based recommenders.
	\item We propose a hierarchical self-attentive recommendation model (AttList), naturally fitting the user-list-item hierarchical structure, and enhanced by a novel self-attentive aggregation layer to capture user and list consistency.
	\item Experiments on three real-world datasets (Goodreads, Spotify, and Zhihu) demonstrate the effectiveness of AttList versus state-of-the-art alternatives. Further, all code and data are released to the research community for further exploration.
\end{itemize}

\section{Related Work}
\label{related work}
%In this section, we highlight related work that informs the design of AttList. % with respect to implicit recommendation, user-generated item lists, and attention networks

%\medskip
\noindent\textbf{Recommendation with Implicit Feedback.}
\label{Item Recommendation with Implicit Feedback}
Early recommender systems mainly focused on rating prediction based on explicit feedback, such as ratings from reviewers \cite{koren2009matrix}. However, implicit feedback is usually much easier to collect, as is the case in our user-list interactions dataset described in Section~\ref{Data Sets}. Thus, more attention has been paid to ranking items based on user preference reflected from implicit feedback, such as purchase history, clicks, or likes. Among the various ranking algorithms \cite{hu2008collaborative, bpr, pan2008one, he2018pseudo}, Bayesian Personalized Ranking (BPR) \cite{bpr} is a well-known pair-wise ranking framework. Recently, there have been some efforts to improve item recommendation by introducing non-linear transformations with neural networks \cite{ncf, cdae, li2017collaborative, liang2018variational, li2015deep, covington2016deep}, where Neural Collaborative Filtering (NCF) \cite{ncf} is a typical example, where a multi-layer perceptron model and a generalized matrix factorization model are combined to learn the user preference. Their experiments show that neural based recommender systems outperform traditional methods like matrix factorization and BPR. 

\smallskip
\noindent\textbf{User-Generated Item Lists.} Recently, user-generated item lists receive more and more interest. Lo et al. \cite{lo2017understanding} analyze the growth of image collections on Pinterest. Lu et al. \cite{liu2016power} and Eksombatchai et al. \cite{eksombatchai2018pixie} distill user preference from user-generated item lists to enhance individual item recommendation. Besides, Greene et al. \cite{greene2011supporting} support users to continue their user-lists on Twitter. In this paper, we focus on recommending user-generated item lists.

%\medskip
%\noindent\textbf{List-Based Recommendation.} 
Although many existing algorithms successfully recommend individual and independent items to users, approaches for recommending user-generated item lists are not fully explored. There are two studies close to ours: LIRE \cite{liu2014recommending} and EFM \cite{cao2017embedding}. The List Recommending Model (LIRE) \cite{liu2014recommending} is a Bayesian-based pairwise ranking approach, which takes user preference over items within the lists into consideration when inferring the user-lists preference, whereby each list is modeled as a linear combination of the items. The Embedding Factorization Model (EFM) \cite{cao2017embedding} uses the co-occurrence information between items and lists to improve list recommendation. These models are based on a \textit{collective matrix factorization framework} \cite{singh2008relational} which jointly factorizes several related matrices, e.g., a user-list interaction matrix and a user-item interaction matrix. In contrast, the proposed AttList is designed to capture the user-list-item hierarchical structure, which is better customized for user-generated item list recommendation. Furthermore, both LIRE and EFM incorporate both user-item and user-list interactions; in contrast, AttList requires only user-list interactions to uncover user preferences though it can be easily extended to incorporate additional user-item interactions as well.%like LIRE and EFM, our model can also easily utilize the interactions between users and extra individual items to improve the performance, which is omitted in this paper for limited space. 

\smallskip
\noindent\textbf{Attention Networks.} AttList incorporates an attention layer inspired by recent work like the attention mechanism proposed in \cite{bahdanau2014neural} and the \textit{transformer} in \cite{vaswani2017attention} for machine translation, where it can be used to relieve the long-range dependency problem in RNNs. Attention networks have been used in recommendation and can be grouped into two classes: vanilla attention and self-attention \cite{zhang2017deep}.

% \textcolor{red}{After that, a novel model \textit{transformer} achieved state-of-the-art performance and efficiency for machine translation\cite{vaswani2017attention}, which uses self-attention to capture the contextual information of each word.} 

\smallskip
\noindent\textit{Vanilla Attention.} In this case, a two-layer network is normally used to calculate the attention score by matching a sequence of representations of the target's components against a learnable global vector. Xiao et al.  \cite{xiao2017attentional} propose an attentional factorization machine model where the importance of feature interactions is learned by attention networks. Zhou et al. \cite{zhou2018micro} employ attention networks to calculate the weights for different user behaviors (e.g., reading the comments, carting and ordering) for modeling the user preference in E-commerce sites. Chen et al.  \cite{chen2017attentive} propose an attentive collaborative filtering framework, where each item is segmented into component-level elements, and attention scores are learned for these components for obtaining a better representation of items. Attention networks are also applied in group recommendation \cite{cao2018attentive}, sequential recommendation \cite{ying2018sequential}, review-based recommendation \cite{Seo2017att, TayMulti-Pointer, chen2018neural} and context-aware recommendation \cite{mei2018attentive}. % Zhai et al. \cite{zhai2016deepintent} improve ads recommendation by weighting words in queries and ads according to their attention scores.

\smallskip
\noindent\textit{Self-Attention.} In this case, attention scores are learned by matching the representations against themselves and update each representation by incorporating the information of other representations. Zhou et al. \cite{zhou2017atrank} apply self-attention networks to capture the inner-relations among different user behaviors for modeling the user preference. Other approaches \cite{kang2018self, zhang2018dynamic, huang2018csan} use self-attention networks for sequential recommendation where the item-item relationship is inferred from user's historical interactions. Self-attention is also applied for point-of-interest recommendation \cite{ma2018point}. In this paper, self-attention is used to distill the consistency of neighboring items and lists for revealing user preferences.% information from lists or user's interactions profile. 

\section{Problem Formulation}
We denote the set of users, lists and items as $\mathcal{U}$ and $\mathcal{L}$ and $\mathcal{T}$ where the size of these sets is  $|\mathcal{U}|$, $|\mathcal{L}|$ and $|\mathcal{T}|$ respectively. Every list in $\mathcal{L}$ is composed of items from $\mathcal{T}$, where the containment relationship between $\mathcal{L}$ and $\mathcal{T}$ is denoted as $\mathcal{C}$. We reserve $u$ to denote a user and $l$ to denote a list. We define the user-list interaction matrix as $\textbf{R} \in \{0,1\}^{|\mathcal{U}| \times |\mathcal{L}|}$, where $r_{ul}$ indicates the feedback from $u$ to $l$, typically in the form of a ``like'', vote,  following, or other feedback signal.  Since the feedback is naturally binary, we further let $r_{ul}=1$ indicate $l$ has feedback from $u$ and $r_{ul}=0$ otherwise. 
% As discussed in Section \ref{Item Recommendation with Implicit Feedback}, implicit feedback is easier to obtain than explicit one and receive more attention recently. We also focus on implicit feedback in the form of behaviors that users like, vote or follow lists.

The \textit{user-generated item list recommendation} problem takes as input the users $\mathcal{U}$, lists $\mathcal{L}$, items $\mathcal{T}$, user-list interactions \textbf{R} and $\mathcal{C}$, the containment relationship between lists $\mathcal{L}$ and items $\mathcal{T}$. It outputs $K$ item lists for each user, where each list $l$ is ranked by the estimated preference of $u$ to $l$, denoted as $\hat{r}_{ul}$. 

\begin{figure*}[htbp]
    \centering
    \includegraphics[scale=0.46]{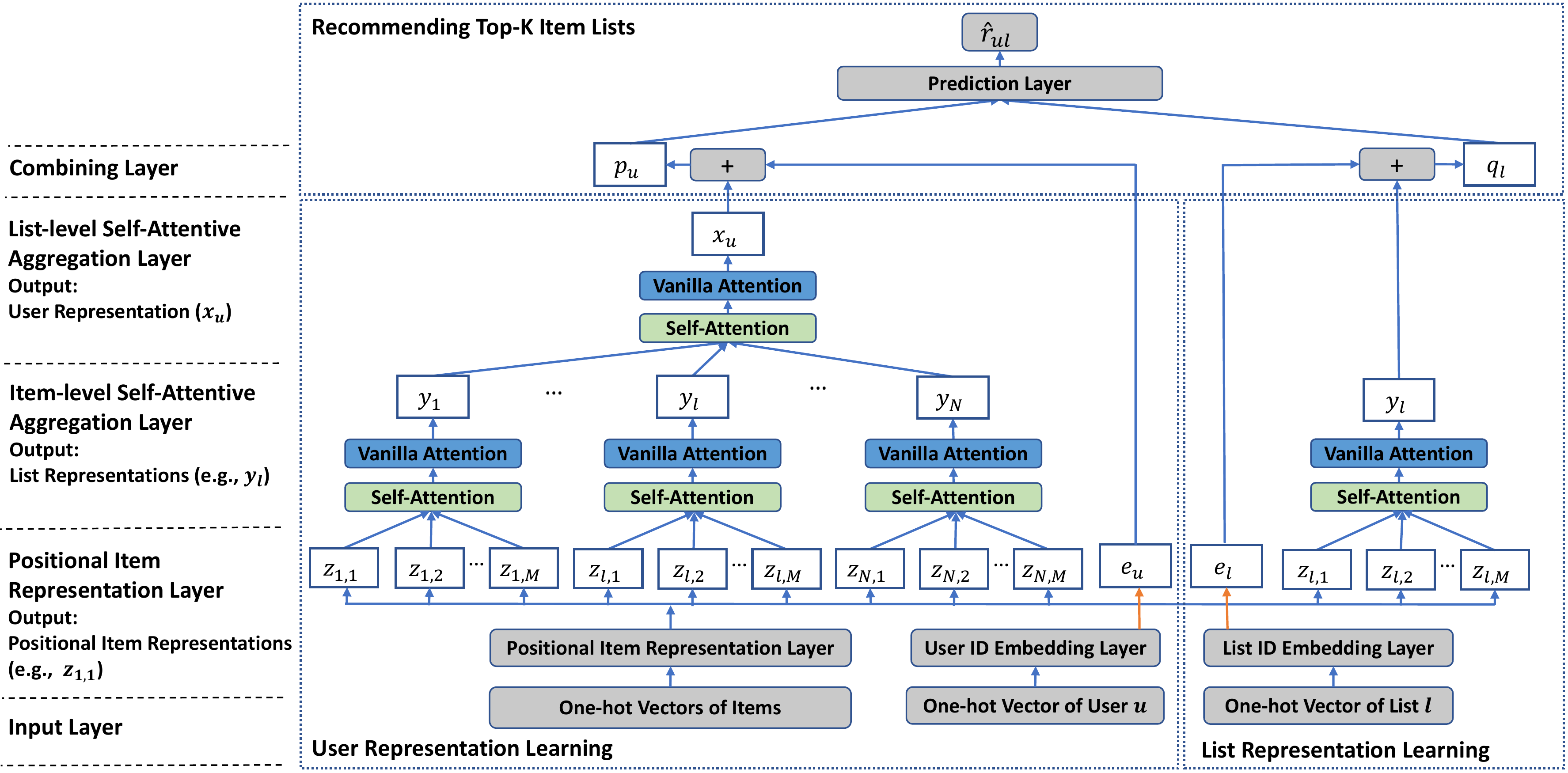}
    \caption{Framework of AttList for recommending user-generated item lists. The model is composed of three components: prediction, user representation learning, and list representation learning. Specifically, the user representation learning component is a stack of four layers: input layer, item positional representation layer, item-level self-attentive aggregation layer, and list-level self-attentive aggregation layer. The list representation learning component also contains the same layers as the user part except for a list-level self-attentive aggregation layer. 
   	}
    \label{model_overview}
\end{figure*}

\section{The Proposed model: AttList}
\label{method}
As discussed in Section \ref{Introduction}, for user-generated item list recommendation, user preference naturally propagates hierarchically from items to lists and finally to users. Furthermore, the user preference on an item is affected by the rest of the items in the same list. Likewise, the user preference for a list is affected by the rest of lists interacted with by the user. Guided by these observations, we propose our hierarchical self-attentive recommendation model (AttList), as shown in Figure \ref{model_overview}, which is built around three research questions:

\begin{itemize}
	\item \textbf{RQ1:} How can we hierarchically model the user preference from items to lists and finally to users?
	\item \textbf{RQ2:} How can we incorporate the consistency of neighboring items and lists to better model user preferences?  
	\item \textbf{RQ3:} Finally, how can we recommend user-generated item lists given the hierarchical preference model?
\end{itemize}

\subsection{RQ1: Hierarchical User Preference Model}
\label{Hierarchical User Preference Model}
The goal of the hierarchical user preference model is to learn the latent representation for a user -- $\textbf{x}_{u}$ -- and the latent representation for a list -- $\textbf{y}_{l}$. Concretely, we propose an item-level aggregation layer to collect user preference on items to represent the list and a list-level aggregation layer to collect user preference on lists to model the user.
 
%\subsubsection{representation Layer}
%\label{representation Layer}

\medskip
\noindent\textbf{Input Layer.} The number of items contained in a list varies and we let $M$ denote the maximum number of items that our model can handle. If a list length is greater than $M$, we only keep the earliest $M$ items curated in the list. If the number of items is less than $M$, a ``padding'' item is repeatedly added to the list until the length is $M$. Likewise, the number of lists interacted with a user varies and we let $N$ denote the maximum number of lists and apply the same padding strategy as items. Thus, the input for the user in a user-list pair is an item ID matrix $\textbf{U} \in \mathbb{N}^{N \times M}$, where each element in $\textbf{U}$ is an item ID. And the input for the list in the user-list pair is an item ID vector $\textbf{L} \in \mathbb{N}^{M}$. Note that $M$ and $N$ are tuned on the validation set and will be discussed in Section \ref{RQ6: Impact of Hyper-parameters}. 

\medskip
\noindent\textbf{Positional Item Representation Layer.} We first create a learnable item representation matrix $\textbf{E} \in \mathbb{R}^{|\mathcal{T}| \times d}$ where $d$ is the latent dimensionality. Based on $\textbf{E}$, the $i$-th item in $l$ can retrieve its representation $\textbf{e}_{li} \in \mathbb{R}^{d}$. For the padding item, a constant zero vector $\textbf{0}$ is used. Following \cite{liu2014recommending}, we also take the position of items in a list into consideration because users usually see the top items in a list first and may stop exploring the item list after enough items have been consumed. Thus, we also create a learnable position representation matrix $\textbf{O} \in \mathbb{R}^{M \times d}$ ($M$ is the length of the list) and $\textbf{o}_{li}$ denotes the position representation for $\textbf{e}_{li}$. We further add them together as the positional item representation:
\begin{equation}
	\textbf{z}_{li} = \textbf{e}_{li} + \textbf{o}_{li}
\end{equation}
The effect of the position representation will be empirically studied in our experiments. 

%\subsubsection{Item-level Pooling Layer} 
%\label{Item-level Pooling Layer}
\medskip
After the item representations are obtained, the next question is how can we aggregate item representations to model the latent representation of the list containing the items:

\medskip
\noindent\textbf{Item-Level Self-Attentive Aggregation Layer.} This layer is composed of two attention networks: a self-attention network \cite{vaswani2017attention} to take consistency of neighboring items to improve the item representations and a vanilla attention network to aggregate the item representations into a list representation space.

We first use a self-attention network to refine the item representations by incorporating the consistency of their neighboring items within a list:

$$
	\textbf{z}_{li}^{\prime} = SelfAttention(\textbf{z}_{li})
$$
where $\textbf{z}_{li}^{\prime} \in \mathbb{R}^{d}$ denotes the representation of the $i$-th item in list $l$, refined by our self-attention network, which will be introduced in detail in Section \ref{Self-attention networks section}.

%\medskip
%\noindent\textbf{Item-level Vanilla Attention network}: 
%Intuitively, the higher the weight is assigned, the more informative the item is to characterize the list.
Then, we apply the weighted sum to aggregate the refined item representations for constructing the list latent representation. The higher the weight is, the more informative the item is to reveal user preference:
\begin{equation}
	\textbf{y}_{l} = \sum_{i \in \mathcal{M}} \alpha_{i} \cdot \textbf{z}_{li}^{\prime}
\end{equation}
where $\textbf{y}_{l} \in \mathbb{R}^{d}$ is the aggregated list representation. And $\alpha_{i}$ is the weight learned by the vanilla attention network:

\begin{equation}
	\begin{split}
	\alpha_{i} = \textbf{u}_{I}^{T} tanh(\textbf{W}_{I} \textbf{z}_{li}^{\prime} + \textbf{b}_{I}), \\
	\alpha_{i} = \frac{exp(\alpha_{i})}{\sum_{n \in \mathcal{M}} exp(\alpha_{n})}, \\
	\end{split}
\end{equation}

\noindent where $\alpha_{i}$ is a scalar as the weight, $\mathcal{M} = \{n \in \mathbb{N}| 1\le n\le M\}$ and $M$ is the length of the list, $\textbf{W}_{I}\in \mathbb{R}^{d \times d}$ is the weight matrix, $\textbf{b}_{I} \in \mathbb{R}^{d}$ is the bias vector, $\textbf{u}_{I} \in \mathbb{R}^{d}$ is the global vector and tanh is used as activation function empirically. The softmax function is used to normalize the weight. Note that the consistency information of the item has been incorporated in $\textbf{z}_{li}$ by the self-attention network, thus $\alpha_{i}$ is learned by considering not only the item's intrinsic properties but also its consistency with neighboring items. 

\medskip
Based on the aggregated list representations, we further aggregate them to model the latent representation of the user who has interacted with the lists:

\medskip
\noindent\textbf{List-Level Self-Attentive Aggregation Layer.} Which is similar to the structure of the item-level aggregation layer, this list-level layer is composed of two attention networks. The first is a self-attention network to improve the list representations by considering the consistency of a user's interaction profile. The second is a vanilla attention network to aggregate the list representations into a user representation space.

%\medskip
%\noindent\textbf{List-level Vanilla Attention network}:

We first refine the list representations by injecting consistency information of their neighboring lists:
$$
	\textbf{y}_{l}^{\prime} = SelfAttention(\textbf{y}_{l})
$$
where $\textbf{y}_{l}^{\prime} \in \mathbb{R}^{d}$ denotes the representation of list $l$, refined by our self-attention network, detailed in Section \ref{Self-attention networks section}.

Then, we select informative lists and aggregate list representations into the user representation space:
\begin{equation}
	\textbf{x}_{u} = \sum_{l \in \mathcal{R}_{u}^{+}} \beta_{l} \cdot \textbf{y}_{l}
\end{equation}
where $\textbf{x}_{u} \in \mathbb{R}^{d}$ is the aggregated user representation. And $\beta_{l}$ is the weight learned by the vanilla attention network, which can be interpreted as the contribution of list $l$ to reveal the user preference:

\begin{equation}
	\begin{split}
	\beta_{l} = \textbf{u}_{L}^{T} tanh(\textbf{W}_{L} \textbf{y}_{l}^{\prime} + \textbf{b}_{L}), \\
	\beta_{l} = \frac{exp(\beta_{l})}{\sum_{l \in \mathcal{R}_{u}^{+}} exp(\beta_{l})}, \\
	\end{split}
\end{equation}

\noindent where $\beta_{l}$ is a scalar as the weight. $\textbf{u}_{L} \in \mathbb{R}^{d}$, $\textbf{W}_{L} \in \mathbb{R}^{d \times d}$, $\textbf{b}_{L} \in \mathbb{R}^{d}$ are the global vector, weight matrix and bias vector respectively. Note that the consistency information of the list has been incorporated in $\textbf{y}_{l}$ by the self-attention network, thus $\beta_{l}$ is learned by considering not only the list's intrinsic properties but also its consistency with neighboring lists. 

In this section, we have proposed a hierarchical user preference model based on the first motivating observation, and next we need to carefully take care of the list and item consistency to better model the user preference based on the second motivating observation.

\subsection{RQ2: Capturing Item and List Consistency by Self-Attentive Networks}
\label{Self-attention networks section}
%\medskip
%\noindent\textbf{Item-level Self-Attention Network} \label{Item Self-Attention network}: 

In this section, a self-attention network is carefully designed to consider the consistency of neighboring items and lists to better model the user preference. Note this self-attention network works in the same way for the item-level aggregation layer and the list-level aggregation layer. For simplicity, we introduce the details for the item-level layer only (the list-level is similarly defined). % and we only introduce the detail of it in the item-level for simplicity. 

%The goal of the self-attention network is to refine the item latent representations by relating each item to other ones within the list.
Item consistency is first measured by the similarities between each item and its neighboring items (in Equation \ref{self-attention equation}) and then preserved in the refined item representations (in Equation \ref{refine equation}) and finally returned to the vanilla attention network for assigning weights to the items (in Section \ref{Hierarchical User Preference Model}). Therefore, not only the item's intrinsic properties but also its ``contextual" items can be considered in the vanilla attention network (in Section \ref{Hierarchical User Preference Model}) to assign weight to the item. Particularly, an item representation will be less injected into the list representation if it is inconsistent with the rest of items in the list.

\begin{figure}
  \centering
   \setlength{\abovecaptionskip}{0.01cm}
  \setlength{\belowcaptionskip}{-0.5cm}
  \subfigure[Self-attention scores $F_{l}$ reflects the similarities among items in list $l$. (Equation \ref{self-attention equation})]{
    \label{selfAttExampleItem} %% label for first subfigure
    \includegraphics[width=1.35in]{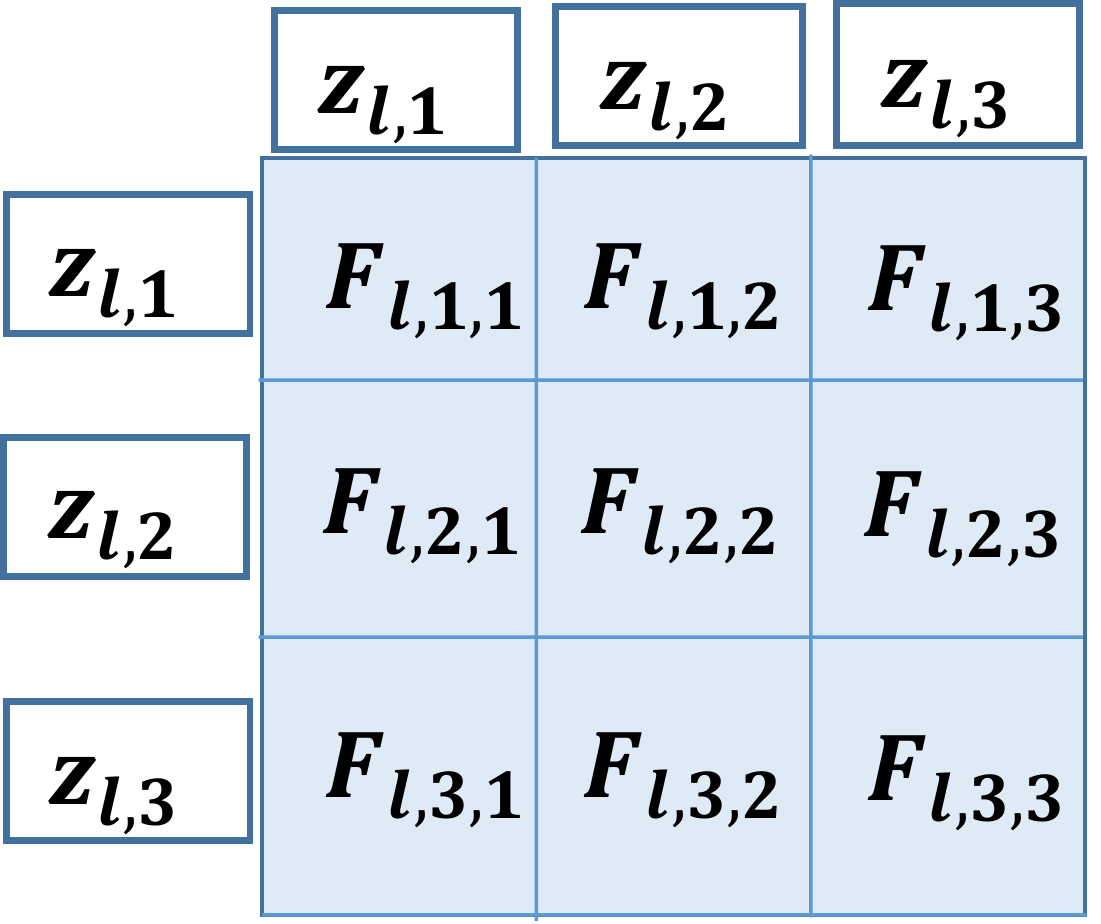}}
  %\hspace{1in}
  \subfigure[Refine item representation by using weighted-sum of original representations, where weights are the self-attention scores. (Equation \ref{refine equation})]{
    \label{selfAttExampleRefine} %% label for second subfigure
    \includegraphics[width=1.6in]{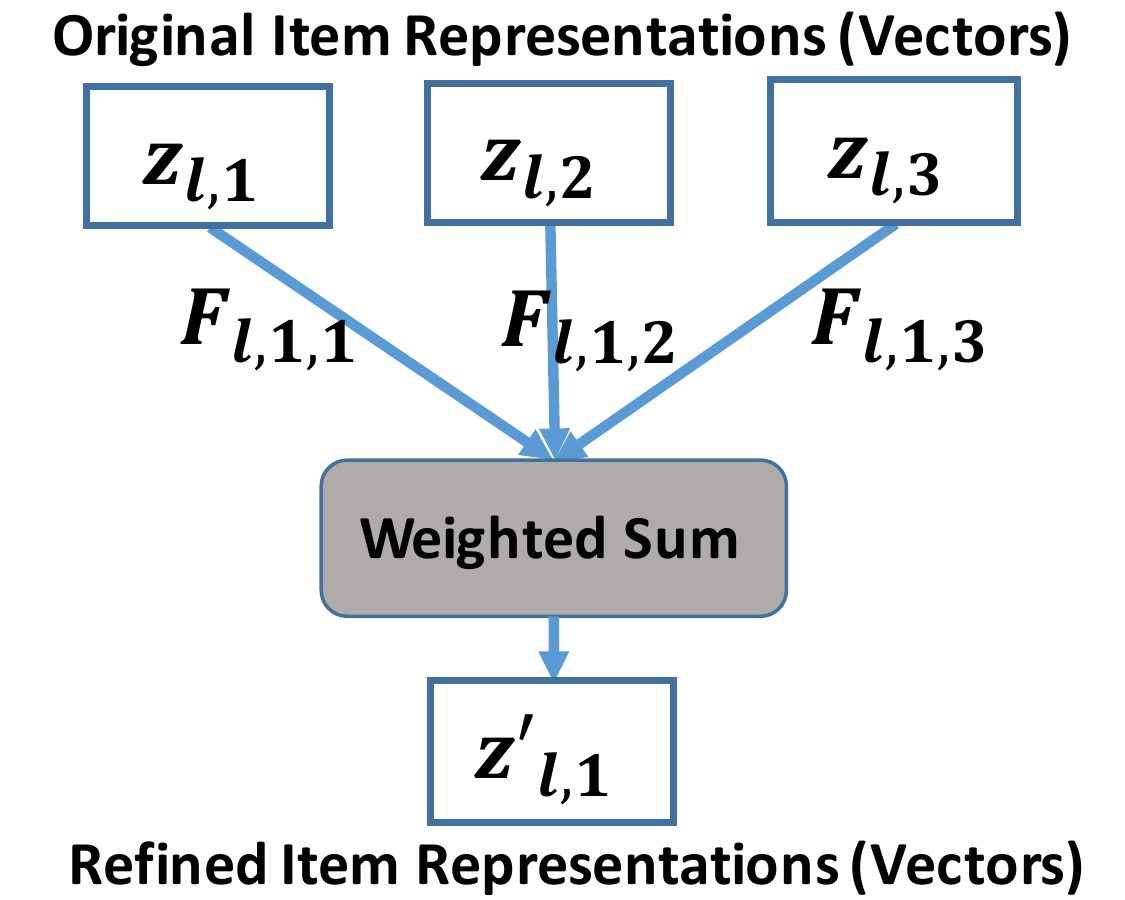}}
   \caption{Refining item representations ($\textbf{z}_{l1}$) by considering the consistency with neighboring items (i.e., $\textbf{z}_{l2}$ and $\textbf{z}_{l3}$).}
  \label{refining item representations} %% label for entire figure
\end{figure}

Specifically, we first pack all item representations from the same list, denoted as $\{\textbf{z}_{li}|i \in \mathbb{N}, 1\le i\le M\}$, together into as a matrix $\textbf{Z}_{l} \in \mathbb{R}^{M \times d}$. Following \cite{vaswani2017attention}, scaled dot-product attention is also used to calculate the self-attention score. We first have:

\begin{equation}
\label{self-attention equation}
	\textbf{F}_{l} = softmax(\frac{\textbf{Z}_{l}\textbf{Z}_{l}^{T}}{\sqrt{d}})
\end{equation}

\noindent where $\textbf{F}_{l} \in \mathbb{R}^{M \times M}$ is a self-attention score matrix which indicates the similarities among $M$ items in a list. Since $d$ is usually a large number (e.g. 128), $\sqrt{d}$ is used to scale the similarity (attention score) for preventing gradient vanishing and the softmax function is to normalize the self-attention scores. An example of Equation \ref{self-attention equation} is illustrated in Figure \ref{selfAttExampleItem}. Then, the output of this network is obtained by multiplying the self-attention score matrix $\textbf{F}_{l}$ with the original item representation matrix $\textbf{Z}_{l}$:

\begin{equation}
    \label{refine equation}
    \textbf{Z}_{l}^{\prime} = \textbf{F}_{l}\textbf{Z}_{l}
\end{equation}

\noindent where $\textbf{Z}_{l}^{\prime} \in \mathbb{R}^{M \times d}$ is a set of updated item representations where each representation is a weighted sum of other ones and the weights are the self-attention scores. In this way, the item representations can be refined by incorporating information of its neighboring items, which is illustrated in Figure \ref{selfAttExampleRefine}. Moreover, to stabilize the training as \cite{vaswani2017attention, kang2018self}, we use residual connections \cite{he2016deep} to obtain the final output:
\begin{equation}
	SelfAttention({Z}_{l}) = \textbf{Z}_{l} + \textbf{Z}_{l}^{\prime}
\end{equation}

\noindent where the original item representation is element-wisely added to the refined item representation. The residual connections benefit the model training by propagating useful low-level features to higher layers. The effect of the residual connections will be discussed in our experiments. Besides, previous work \cite{vaswani2017attention, kang2018self} has shown how to first convert the input representation matrix into three matrices through three linear projections and then feed them to the scaled dot-product attention. Though this way leads to a more expressive model, it increases the training difficulty and harms the performance in our experiments. So far, we have improved the user preference model by considering the item and list consistency, and next the model (i.e., $x_{u}$, $y_{l}$) is utilized to recommend item lists.

\subsection{RQ3: Recommending Top-K Item Lists}
\label{Recommending Top-K Item Lists}
Finally, we are ready to recommend the top-k item lists. We first combine the aggregated representations with the ID embedding of items and lists to capture general user preferences and then design a prediction layer to calculate the preference score for recommending item lists. 

\medskip
\noindent\textbf{Combining Layer}.\label{Combining Layer} In this layer, we element-wisely add the aggregated user representation ($\textbf{x}_{u}$) with a dedicated ID embedding of the user ($\textbf{e}_{u} \in \mathbb{R}^{d}$):
\begin{equation}
	\textbf{p}_{u} = \textbf{x}_{u} + \textbf{e}_{u}
\end{equation} 

\noindent where $\textbf{p}_{u} \in \mathbb{R}^{d}$ is the final representation of user $u$. The intention is to consider the general user preference (modeled by $ \textbf{e}_{u}$) that in some cases cannot be fully characterized by the aggregation of the list representations, which has been proven beneficial for recommendation in \cite{chen2017attentive, cao2018attentive, chen2018neural}. Similarly, we have:

\begin{equation}
	\textbf{q}_{l} = \textbf{y}_{l} + \textbf{e}_{l}
\end{equation}

\noindent where $\textbf{y}_{l} \in \mathbb{R}^{d}$ is the aggregated representation of $l$ of list $l$, $\textbf{q}_{l} \in \mathbb{R}^{d}$ is the final representation and a dedicated ID embedding of the list $\textbf{e}_{l} \in \mathbb{R}^{d}$ is also learned to model the general characteristic of the list, which cannot be fully expressed by the aggregation of the item representations.% as discussed in Section \ref{Combining Layer}.

\medskip
\noindent\textbf{Prediction Layer}. In this layer, the estimated preference score $\hat{r}_{ul}$ is calculated given the user latent representation $\textbf{p}_{u}$ and the list latent representation $\textbf{q}_{l}$. Following \cite{ncf, cao2018attentive}, we also first apply the element-wise product on $\textbf{p}_{u}$ and $\textbf{q}_{l}$, and then concatenate  it with the original representations:
\begin{equation}
	\textbf{h}_{0} = 
	\begin{bmatrix}
		\textbf{p}_{u} \odot \textbf{q}_{l} \\ \textbf{q}_{l} \\ \textbf{p}_{u}
	\end{bmatrix}
\end{equation}

\noindent where the element-wise product $\textbf{p}_{u} \odot \textbf{q}_{l}$ captures the interactive relationship between the user and the list which follows traditional latent factor models. The concatenation of the original representations $\textbf{p}_{u}$ and $\textbf{q}_{l}$ is to prevent information loss due to the element-wise product which has been proven effective for recommendation in \cite{ncf, cao2018attentive}.

After that, we apply a two-layer neural network to obtain the final preference score:
\begin{equation}
\left\{
             \begin{array}{lr}
             \textbf{h}_{1} = ReLU(\textbf{W}_{1}\textbf{h}_{0} + \textbf{b}_{1}), &  \\
             %y=s, & 0\leq s\leq L,|t|\leq1.\\
             \hat{r}_{ul} = Sigmoid(\textbf{W}_{2}\textbf{h}_{1} + \textbf{b}_{2}) &  
             \end{array}
\right.
\end{equation}

\noindent where $\textbf{h}_{1} \in \mathbb{R}^{D}$ is the output of the hidden layer, $\textbf{W}_{1} \in \mathbb{R}^{D \times 3d}$ and $\textbf{W}_{2} \in \mathbb{R}^{1 \times D}$ denote the weight matrix, $\textbf{b}_{1} \in \mathbb{R}^{D}$ and $\textbf{b}_{2} \in \mathbb{R}^{D}$ denote the bias vector, in which $D$ as the predictive factors \cite{ncf} controls the model capability. Empirically, ReLU function is used as the non-linear activation function for the hidden layer. For each user, all lists are ranked by their corresponding preference scores and the top-k lists are recommended to the user.

\medskip
\noindent\textbf{Objective Function.} Following \cite{ncf}, we also treat top-K recommendation with implicit feedback as a binary classification problem and apply the binary cross-entropy loss as the loss function:

\begin{equation}
	\ell = - \sum_{u \in \mathcal{U}}\sum_{\mathcal{R}_{u}^{+} \cup \mathcal{R}_{u}^{-}} r_{ul} \, log \, \hat{r}_{ul} + (1-r_{ul}) \, log \, (1-\hat{r}_{ul})
\end{equation} 

\noindent where $\mathcal{R}_{u}^{+}$ denotes the set of lists interacted with $u$ and $\mathcal{R}_{u}^{-}$ is the set of negative samples which are uniformly sampled from unobserved interactions ($r_{ul}=0$). Note that we generate a new batch of negative samples in each iteration and the amount of the negative samples can be controlled by a hyper-parameter ratio $\rho$, i.e., $|\mathcal{R}_{u}^{-}| = \rho|\mathcal{R}_{u}^{+}|$.

\section{Experimental Setup}
In this section, we first introduce three list-based datasets. After that, baseline methods, details for reproducibility and evaluation metrics are also provided.

\subsection{Datasets: Goodreads, Spotify, and Zhihu}
\label{Data Sets}
In this section, we describe our three list-based datasets, which are are drawn from three popular platforms with different kinds of lists (book-based, song-based, and answer-based), reflecting a variety of scenarios to evaluate AttList and alternative approaches. Table \ref{data_statistics} summarizes these three datasets. %three user generated lists datasets to evaluate our proposed approach. The scenarios of consuming user generated lists and the our data crawling strategies will be introduced.

\medskip
\noindent\textbf{Goodreads.} Goodreads is a popular site for book reviews and recommendation. Users in Goodreads can browse all the books within a book list and vote for the list to express their preference. We randomly sample 18,435 users from all the users and then crawl all lists voted by them as well as the books within the lists, resulting in 24,217 lists containing 158,392 books (items).

%Since the user identification numbers in Goodreads are positive integers starting from 1, it is non-trivial to randomly sample the users and formulate the urls for their profiles to be crawled. However, for the other two datasets, identification numbers have no such rule and hard to be sampled, Hence, we apply different sampling and crawling strategies according to their APIs and usages.

\medskip
\noindent\textbf{Spotify.} Spotify is a music streaming service where users can follow playlists created by other users. We first search for playlists by issuing 200 keyword queries representing popular topics on Spotify, like ``pop'' and ``drive''. Then, followers of the playlists are further crawled, arriving at 7,787 lists containing 49,434 songs (items).

\medskip
\noindent\textbf{Zhihu.} Zhihu is a question and answer community where users can follow lists of answers that have been curated by other users. For example, a list may contain two answers as the response to two questions correspondingly: ``What is the best way to read a book?" and ``How to dress for an interview?". We first randomly sample users from the answerers of three popular topics: life, learning and recreation. Then the lists followed by the users are also crawled, resulting in 12,715 lists containing 211,242 answers (items).

\medskip
Following the preprocessing setup in \cite{liu2014recommending, cao2017embedding}, we filter out items appearing in fewer than 5 lists.  To study the impact of data density on recommendation performance, we also filter out users who have interacted with lists fewer than 5 times on Spotify and Zhihu, so that they are denser than the Goodreads dataset. Even with such filtering, note that the three datasets are all very sparse with density of 0.056\%, 0.115\% and 0.178\% respectively, which reflects the data sparsity challenge in real-world list recommendation. 

\begin{figure}
  \centering
     \setlength{\abovecaptionskip}{0.1cm}
  \setlength{\belowcaptionskip}{-0.6cm}
  \subfigure[User-list Interaction Relationship]{
    \label{User-list} %% label for first subfigure
    \includegraphics[width=1.5in]{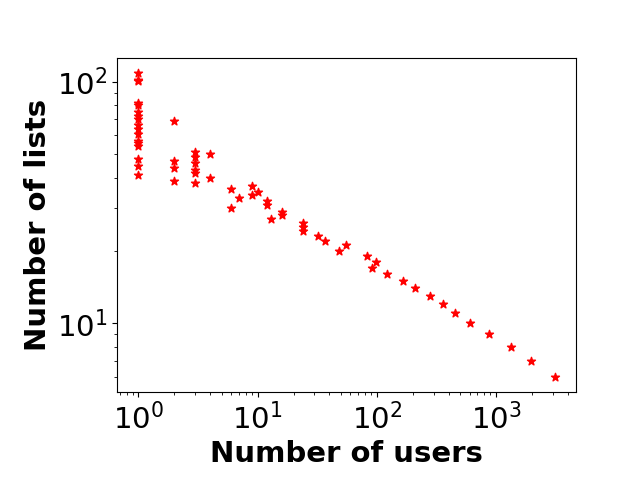}}
  %\hspace{1in}
  \subfigure[List-item Containing Relationship]{
    \label{List-item} %% label for second subfigure
    \includegraphics[width=1.5in]{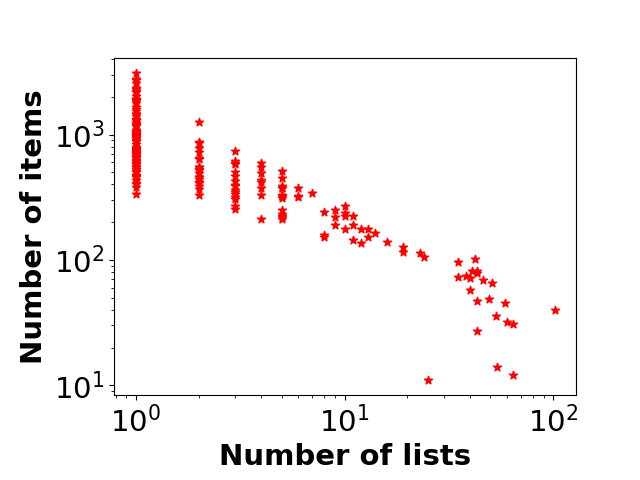}}
   \caption{Power law distribution between users \& lists and lists \& items in Spotify dataset.}
  \label{Power law distribution} %% label for entire figure
\end{figure}

%{\color{red}{Then, we also implement a statistical analysis and show the results of the preprocessed Spotify dataset as the example. }} 
%Unsurprisingly, we find that most users interact with very few lists, as our data sparsity hints at. 
In Figure \ref{User-list}, we group users into buckets according to how many lists they interacted with and show the number of lists that belong to the buckets correspondingly. We find that most users only interact with a few lists while a few users interact with a large number of lists. A similar observation can be seen in Figure \ref{List-item}, where most lists contain a few items while a few lists contain a large number of items. Thus, we observe a power law distribution between the number of lists a user interacts and the number of items a list contains.

%Different from \cite{liu2014recommending, cao2017representation}, we keep all the users and lists if the users interact with at least one list and the lists contain at least one item. Thus, the three datasets are extremely sparse 

For our experiments, we randomly split each dataset into three parts: 80\% for training, 10\% for validation and 10\% for testing. Since many users have very few interactions, they may not appear in the testing set at all. In this case, there will be no match between the top-K recommendations of any algorithm and the ground truth since there are no further interactions made by these users. The results reported here include these users, reflecting the challenges of list recommendation under extreme sparsity.

\begin{table}[htbp]
  \centering
  %\small
       \setlength{\abovecaptionskip}{0.01cm}
  \setlength{\belowcaptionskip}{-0.3cm}
  \renewcommand{\arraystretch}{0.8}
	\setlength{\tabcolsep}{2.1pt}
  \caption{Summary Statistics for the Evaluation Datasets}
    \begin{tabular}{|l|c|c|c|c|c|}
	\toprule
    Dataset & \#Users & \#Lists & \#Interactions & Density  & \#Unique Items \\ 
	\midrule
    Goodreads & 18,435 & 24,217 & 250,450 & 0.056\% & 158,392 \\
    Spotify & 10,183 & 7,787 & 91,254 & 0.115\% & 49,434 \\
    Zhihu & 6,101 & 12,715 & 138,458 & 0.178\% & 211,242 \\
	\bottomrule
    \end{tabular}%
  \label{data_statistics}%
\end{table}%

\subsection{Baselines}
In this section, we introduce a suite of baselines. The first set of methods are item-based top-k recommendation methods that we adapt to the problem of list recommendation:

\medskip
\noindent\textbf{ItemPop.} This simple method recommends the most popular item lists. Lists are ranked by the number of times they have been interacted with (e.g., followed or liked). The same top-K most popular lists are recommended to all users. 

\medskip
\noindent\textbf{MF.} This is the traditional matrix factorization (MF) method proposed in \cite{koren2009matrix} using mean squared error as the objective function with negative sampling from the non-observed lists ($r_{ul}=0$). 

\medskip
\noindent\textbf{BPR.} Bayesian personalized ranking (BPR) \cite{bpr} is a well-known pair-wise ranking framework for implicit recommendation. BPR assumes that, for each user, the preference of the observed list ($r_{ul}=1$) is superior to the non-observed ones ($r_{ul}=0$). 

\medskip
\noindent\textbf{NCF.} Neural collaborative filtering (NCF) \cite{ncf} is a state-of-the-art neural network method for recommendation. NCF concatenates latent factors learned from a generalized matrix factorization model and a multi-layered perceptron model and then uses a regression layer to predict the preferences of a user to the lists. Following \cite{ncf}, negative samples are uniformly sampled from unobserved interactions ($r_{ul}=0$). %The number of negative samples is tuned as a hyper-parameter over the validation set.

\medskip
The next set of methods is designed specifically for list recommendation. They use user-list interactions as well as the containing relationship between lists and items.

\medskip
\noindent\textbf{LIRE.} List Recommendation Model (LIRE) \cite{liu2014recommending} is a Bayesian-based pair-wise ranking approach. It models user preferences by the sum of two factors. The first factor is the inner product between user latent factors and list latent factors; the second factor is the sum of inner products between the user factor and item latent factors within the list. 

\medskip
\noindent\textbf{EFM.} Embedding Factorization Model (EFM) \cite{cao2017embedding} is also a Bayesian-based pair-wise model designed for list recommendation. Inspired by paragraph2vec model in \cite{le2014distributed}, it calculates the shifted positive pointwise mutual information (SPPMI) value between lists and items and uses this information to boost the ranking performance.

\subsection{Reproducibility}
All code and data are available at here\footnote{https://github.com/heyunh2015/AttList}. We implement ItemPop, BPR, and MF. The implementation of NCF is from the authors.\footnote{https://github.com/hexiangnan/neural\_collaborative\_filtering} For LIRE and EFM, we use code from the authors of EFM.\footnote{https://listrec.wixsite.com/efms} We implement AttList with Keras, and Adam \cite{kingma2014adam} is applied as the optimizer. For AttList and baseline methods, all hyper-parameters are tuned on the validation dataset, where early stopping strategy is applied such that we terminate training if validation performance does not improve over 10 iterations. All neural network models were trained using Nvidia GeForce GTX Titan X GPU with 12 GB memory and 3,072 cores.%More details for reproducing our experiments like specific parameter settings can be found in Appendix \ref{extra}. %For all baselines, we tune hyper-parameters by using the validation set and apply early stopping strategy where we terminate training if validation performance does not improve over 10 iterations

%\medskip
%\noindent\textbf{Unshared item-level self-attentive aggregation layer for Lists.} Each list in AttList is modeled in the aggregation of the user and also as the individual list interacted with by the user. For example, $\textbf{y}_{l}$ appears both in the user representation learning component and the list representation learning component, as shown in Figure \ref{model_overview}. We apply two different item-level self-attentive aggregation layers respectively for these two situations. Our consideration is that the list representation serving as a component of the user representation has different characteristics when it interacts with the user. 

\medskip
\noindent\textbf{Eliminating Lists from the Testing Dataset for User Representation.} As discussed in Section \ref{Hierarchical User Preference Model}, AttList represents a user by the aggregation of list representations. In the training, it is important to eliminate the lists which are sampled into the testing dataset to represent the user, which makes the training and testing datasets mutually exclusive. %Otherwise, a false very promising performance will be obtained.

\medskip
\noindent\textbf{Over-fitting Prevention.} Dropout \cite{srivastava2014dropout} is widely used in neural networks training to improve generalization performance. We also apply dropout in every layer of the model. Specifically, we randomly drop $\gamma$ percentage of the output vectors: $\textbf{e}_{li}$, $\textbf{o}_{li}$, $\textbf{y}_{l}$, $\textbf{x}_{u}$ and $\textbf{h}_{1}$. Following \cite{vaswani2017attention}, we also drop $\gamma$ percent of the output of the softmax function in the self-attention network. We also apply $L_{2}$ regularization to the weight matrix in the vanilla attention network (e.g., $\textbf{W}_{I}$ and $\textbf{W}_{L}$), where $\lambda$ denotes the parameter controlling the regularization strength.

\medskip
\noindent\textbf{Parameter Settings.} The batch size is tested from [16, 32, 64, 128, 256] and 32 is selected for all three datasets according to the results on the validation dataset. The learning rate is tested from [0.00005, 0.0001, 0.0005, 0.001, 0.005] and 0.0001 is selected for Spotify and Zhihu while 0.001 is better for Goodreads. The candidates for the latent dimensionality $d$ is from [8, 16, 32, 64, 96] and we select 64 for Zhihu and 96 for the other two datasets. The maximum number of lists $N$ and the maximum number of items $M$ are selected from [5, 10, 15, 20] and [8, 16, 32, 64] respectively. The validation results show that $N=15$ and $M=32$ are better for performance. The ratio for negative sampling $\rho$ is tested from [3, 5, 7, 9, 11] and 3, 5, 7 are selected for Goodreads, Spotify and Zhihu respectively. The predictive factor $D$ is tested from [20, 50, 100] and set as 100 for all datasets. In our model, many components use dropout and L2 regularization. For simplicity, we introduce the range for their parameters tuning: the $\gamma$ percentage for dropout is tested from [0, 0.3, 0.5, 0.8] and the L2 regularization strength $\lambda$ is selected from [0.001, 0.01, 0.1]. 

\subsection{Evaluation Metrics}
Given a user, a top-K item list recommendation algorithm provides a list of ranked item lists according to the predicted preference of them. To assess the ranked lists with respect to the ground-truth lists set of what users actually interacted with, we adopt three evaluation metrics: Normalized Discounted Cumulative Gain (NDCG) \cite{jarvelin2002cumulated} at 5 and 10 (N@5 and N@10), precision at 5 and 10 (P@5 and P@10), and recall at 5 and 10 (R@5 and R@10).

%including precision at 5 (P@5), 10 (P@10), recall at 5 (R@5), 10 (R@10) and F1 value at 5 (F1@5), 10 (F1@10). 

%Precision@K measures the fraction of correctly predicted items among the top-K recommended items:
%
%\begin{equation}
%P@K=\sum_{i=1}^{K}\frac{R(K)}{K}
%\end{equation}
%where $R(i)$ is an indication function of the \textit{i}-th item, $R(i)=1$ indicates the \textit{i}-th item is purchased by the user and 0 otherwise.
%
%Recall measures the fraction of correctly predicted items among all the items purchased by the user:
%
%\begin{equation}
%R@K=\sum_{i=1}^{K}\frac{R(K)}{N(u)}
%\end{equation}
%where $N(u)$ is the total number of items purchased by user $u$ in the testing dataset.
%
%The F1 value is the harmonic mean of precision and recall:
%
%\begin{equation}
%F1@K=\frac{2 \cdot P@K \cdot R@K}{P@K+R@K}
%\end{equation}

\section{Experimental Results and Analysis}
In this section, we present our experimental results and discussion toward answering the following experimental research questions (RQs):

\begin{itemize}
	\item \textbf{RQ4:} How well does AttList perform compared to traditional top-k recommenders? And more importantly, how well does AttList perform compared to models that are specifically designed for list recommendation (LIRE and EFM)?%model outperforms baseline methods especially LIRE and EFM which are designed for list recommendation?
	\item \textbf{RQ5:} What is the impact of the design choices of AttList on the quality of list recommendation? Is attention important?  %various components in the AttList architecture?
	%\item \textbf{RQ3} - What are the influences of the key hyper-parameters on model performance?
	\item \textbf{RQ6:} What is the impact of the hyper-parameters of AttList on the quality of list recommendation? 
	\item \textbf{RQ7:} Finally, do the learned self-attention scores capture meaningful patterns (e.g., consistency) of items and lists?
\end{itemize}

\subsection{RQ4: List Recommendation Quality}
Tables \ref{Results on Goodreads Dataset}, \ref{Results on Spotify Dataset} and \ref{Results on Zhihu Dataset} report the experimental results on three datasets, where `**' indicates that the improvements over all baselines pass the t-test significance test with p-value < 0.01 and particularly $\triangle_{EFM}$ and $\triangle_{LIRE}$ denote the improvement upon EFM and LIRE. We observe that our model significantly outperforms all baselines on all metrics. For instance, AttList achieves the best NDCG@10 of 3.691\%, 8.953\% and 3.666\% on the three datasets, with an improvement upon LIRE of 10.4\%, 5.9\% and 6.8\%. Significant improvements in terms of precision and recall are also observed.

We also observe that the list recommending methods (AttList, LIRE, EFM) are consistently superior to the traditional top-k recommendation methods (ItemPop, BPR, MF, NCF). The possible reason is that the list recommending methods learn the user preference not only from user-list interactions but also from the information of items within the lists. This demonstrates that the top-k item list recommendation is quite different from the general top-k items recommendation and requires considering the containing relationship between lists and items. Besides, all three list recommending methods achieve a better performance in Spotify and Zhihu than in Goodreads, which shows that the denser the dataset is, the higher performance can be obtained by the list recommending methods.

% Table generated by Excel2LaTeX from sheet 'goodreads_latex'
\begin{table}[htbp]
  \centering
   \setlength{\abovecaptionskip}{0.1cm}
  \setlength{\belowcaptionskip}{-0.3cm}
  \setlength{\tabcolsep}{2.0pt}
  \renewcommand\arraystretch{0.8}
  \caption{Experimental Results on Goodreads Dataset}
    \begin{tabular}{|l|cccccc|}
    \toprule
     Metric(\%)  & \multicolumn{1}{l}{P@5} & \multicolumn{1}{l}{R@5} & \multicolumn{1}{l}{N@5} & \multicolumn{1}{l}{P@10} & \multicolumn{1}{l}{R@10} & \multicolumn{1}{l|}{N@10} \\
    \midrule
    ItemPop & 0.692 & 1.832 & 1.487 & 0.538 & 2.620 & 1.760 \\
    MF    & 0.883 & 2.128 & 1.708 & 0.765 & 3.417 & 2.165 \\
    BPR   & 1.004 & 2.270 & 2.043 & 0.852 & 3.801 & 2.597 \\
    NCF   & 1.086 & 2.668 & 2.192 & 0.900 & 4.167 & 2.694 \\
    EFM   & 1.311 & 3.347 & 2.587 & 1.045 & 5.027 &  3.214\\
    LIRE  & 1.337 & 3.271 & 2.731 & 1.068 & 4.965 & 3.345 \\
    \midrule
    AttList & $1.449^{**}$ & $3.580^{**}$ & $3.048^{**}$ & $1.148^{**}$ & $5.465^{**}$ & $3.691^{**}$ \\
    $\triangle_{EFM}$ & 10.6\% & 6.9\% & 17.8\% & 9.9\% & 8.7\% & 14.8\% \\
    $\triangle_{LIRE}$ & 8.4\% & 9.54\% & 11.6\% & 7.6\% & 10.1\% & 10.4\% \\
    \bottomrule
    \end{tabular}%
  \label{Results on Goodreads Dataset}%
\end{table}%

% Table generated by Excel2LaTeX from sheet 'spotify5_2'
\begin{table}[htbp]
  \centering
  \setlength{\abovecaptionskip}{0.1cm}
  \setlength{\belowcaptionskip}{-0.3cm}
  \setlength{\tabcolsep}{1.6pt}
  \renewcommand\arraystretch{0.8}
  \caption{Experimental Results on Spotify Dataset}
    \begin{tabular}{|l|cccccc|}
    \toprule
     Metric(\%)  & \multicolumn{1}{l}{P@5} & \multicolumn{1}{l}{R@5} & \multicolumn{1}{l}{N@5} & \multicolumn{1}{l}{P@10} & \multicolumn{1}{l}{R@10} & \multicolumn{1}{l|}{N@10} \\
    \midrule
    ItemPop & 0.621 & 2.113 & 1.454 & 0.532 & 3.533 & 1.961 \\
    MF    & 2.396 & 7.884 & 6.005 & 1.609 & 10.385 & 6.966 \\
    BPR   & 2.555 & 8.541 & 6.183 & 1.775 & 11.673 & 7.389 \\
    NCF   & 2.602 & 8.793 & 6.643 & 1.820 & 11.914 & 7.832 \\
    EFM   & 2.744 & 9.085 & 6.934 & 1.989 & 13.079 & 8.305 \\
    LIRE  & 2.850 & 9.468 & 7.065 & 2.020 & 13.277 & 8.453 \\
    \midrule
    AttList & $2.944^{**}$ & $9.887^{**}$ & $7.520^{**}$ & $2.110^{**}$ & $13.935^{**}$ & $8.953^{**}$ \\
    $\triangle_{EFM}$    & 7.3\% & 8.8\% & 8.5\% & 6.1\% & 6.6\% & 7.8\% \\
    $\triangle_{LIRE}$    & 3.3\% & 4.4\% & 6.4\% & 4.5\% & 5.0\% & 5.9\% \\
    \bottomrule
    \end{tabular}%
  \label{Results on Spotify Dataset}%
\end{table}%

% Table generated by Excel2LaTeX from sheet 'zhihuLarge5'
\begin{table}[htbp]
  \centering
  \setlength{\abovecaptionskip}{0.1cm}
  \setlength{\belowcaptionskip}{-0.3cm}
  \setlength{\tabcolsep}{2.0pt}
  \renewcommand\arraystretch{0.8}
  \caption{Experimental Results on Zhihu Dataset}
    \begin{tabular}{|l|cccccc|}
    \toprule
      Metric(\%) & \multicolumn{1}{l}{P@5} & \multicolumn{1}{l}{R@5} & \multicolumn{1}{l}{N@5} & \multicolumn{1}{l}{P@10} & \multicolumn{1}{l}{R@10} & \multicolumn{1}{l|}{N@10} \\
    \midrule
    ItemPop & 1.138 & 2.275 & 1.860 & 0.959 & 3.951 & 2.419 \\
    MF & 1.341 & 2.706 & 2.163 & 1.059 & 4.052 & 2.629 \\
    BPR & 1.347 & 2.743 & 2.381 & 1.162 & 4.545 & 3.027 \\
    NCF & 1.485 & 2.834 & 2.435 & 1.216 & 4.674 & 3.057 \\
    EFM & 1.508 & 3.267 & 2.718 & 1.262 & 5.341 & 3.358 \\
    LIRE & 1.596 & 3.319 & 2.853 & 1.303 & 5.350 & 3.433 \\
    \midrule
    AttList & $1.754^{**}$ & $3.600^{**}$ & $2.986^{**}$ & $1.393^{**}$ & $5.646^{**}$ & $3.666^{**}$ \\
     $\triangle_{EFM}$   & 16.3\% & 10.2\% & 9.9\% & 10.4\% & 5.7\% & 9.2\% \\
     $\triangle_{LIRE}$     & 9.9\% & 8.5\% & 4.7\% & 6.9\% & 5.5\% & 6.8\% \\
    \bottomrule
    \end{tabular}%
  \label{Results on Zhihu Dataset}%
\end{table}%

\subsection{RQ5: Ablation Analysis}
% Table generated by Excel2LaTeX from sheet 'ablation'
\begin{table*}[htbp]
  \centering
  \setlength{\abovecaptionskip}{0.1cm}
  \setlength{\belowcaptionskip}{-0.5cm}
	\setlength{\tabcolsep}{3.5pt}
  \renewcommand\arraystretch{0.8}
  \caption{Ablation analysis on the three datasets for precision and recall. Qualitatively similar results hold for NDCG as well. }
    \begin{tabular}{|l|cccc|cccc|cccc|}
	\toprule
          & \multicolumn{4}{c|}{Goodreads} & \multicolumn{4}{c|}{Spotify}   & \multicolumn{4}{c|}{Zhihu} \\
	\midrule
    Architecture & \multicolumn{1}{c}{P@10} & \multicolumn{1}{c}{Change} & \multicolumn{1}{c}{R@10} & \multicolumn{1}{c|}{Change} & \multicolumn{1}{c}{P@10} & Change & \multicolumn{1}{c}{R@10} & Change & \multicolumn{1}{c}{P@10} & Change & \multicolumn{1}{c}{R@10} & Change \\
	\midrule
    AttList & \textbf{1.148} &       & \textbf{5.465} &       & \textbf{2.110} &       & \textbf{13.935} &       & \textbf{1.393} &       & 5.646 &  \\
    - Vanilla Attention & 1.075 & -6.4\% & 4.917 & -10.0\% & 1.756 & -16.8\% & 11.477 & -17.6\% & 1.315 & -5.6\% & 5.347 & -5.3\% \\
    - Self-Attention & 1.111 & -3.2\% & 5.355 & -2.0\% & 1.972 & -6.6\% & 12.988 & -6.8\% & 1.292 & -7.3\% & 5.117 & -9.4\% \\
    - Attention Mechanism &    0.994   &   -13.4\%    &   4.728    &   -13.5\%    & 1.741 & -17.5\% & 11.429 & -18.0\% & 1.252 & -10.1\% & 5.009 & -11.3\% \\
    - Residual Connections &   1.020    &   -11.2\%    &   4.752    &   -13.1\%    & 1.959 & -7.2\% & 12.999 & -6.7\% & 1.254 & -10.0\% & 4.939 & -12.5\% \\
    - Position Information &   1.132    &   -1.4\%    &   5.456    &   -0.2\%    & 2.108 & -0.1\% & 13.904 & -0.2\% & 1.365 & -2.0\% & \textbf{5.668} & +0.4\% \\
    - ID Embedding &    1.044   &   -9.1\%    &   5.114    &   -6.4\%    & 2.093 & -0.8\% & 13.797 & -1.0\% & 1.336 & -4.1\% & 5.571 & -1.3\% \\
    %Remove PT & 1.142 & -0.6\% & \textbf{5.548} & +1.5\% & \textbf{2.145} & +1.6\% & \textbf{14.117} & +1.3\% & 1.364 & -2.1\% & 5.574 & -1.3\% \\
    + Linear Projections & 1.115 & -2.9\% & 5.360 & -1.9\% & 2.027 & -4.0\% & 13.325 & -4.4\% & 1.306 & -6.2\% & 5.213 & -7.7\% \\
    \bottomrule
	\end{tabular}%
  \label{Ablation analysis}%
\end{table*}%

Since AttList is composed of several important design decisions -- like the inclusion of self-attention -- we next present an ablation analysis to study the impacts of these decisions on recommendation quality. %and answer \textbf{RQ2}. 
In Table \ref{Ablation analysis}, we present the results of AttList versus several variants as described next:% of our default model and its variants. The impacts of the components are analyzed respectively as follow:

%\begin{description}
%\begin{itemize}
\smallskip
\noindent\textbf{-- Vanilla Attention:} In this first experiment, all the vanilla attention networks in the architecture are replaced with a simple average pooling method. We observe that performance becomes worse in all three datasets with a large drop, for example, 16.8\% in P@10 for Spotify. This shows that different items and lists reveal fundamentally different evidence of user preferences. This demonstrates that AttList is able to select informative items and lists to characterize the list and the user respectively.

\smallskip
\noindent\textbf{-- Self-Attention:} In this experiment, all the self-attention networks are removed. Significant drops in all metrics are observed across all three datasets, especially in Zhihu with a drop of 7.3\% in P@10. These results show that our self-attention networks can boost performance by considering consistency of neighboring items and lists to refine the item and list representations.

\smallskip
\noindent\textbf{-- Attention Mechanism:} We next not only remove the self attention networks (\textit{-- Self-Attention}) but also replace the vanilla attention networks with the average pooling method  (\textit{-- Vanilla Attention}).  In other words, the arithmetic mean of item representations is used to model the list representation and list representations are just averaged to represent the user in this case. Unsurprisingly, by entirely removing the attention mechanism, we see a larger drop than in either of the previous two isolated experiments. 

\smallskip
\noindent\textbf{-- Residual Connections:} The impacts of removing residual connections in the self-attention networks is quite easily noticed, leading to a large performance drop on all datasets, proving that residual connections are very useful to stabilize the neural network training.

\smallskip
\noindent\textbf{-- Position Information:} The experimental results show that the positional representation has a minor impact on our architecture where a drop of 1.4\%, 0.1\% and 2.0\% is observed on the three datasets respectively. A possible reason is that the maximum number of items that our model can handle is set as 32 according to the hyper-parameters tuning in the validation set. Hence, the influence of the top 32 items does not vary too much in terms of their positions.

\smallskip
\noindent\textbf{-- User and List ID embeddings:} We observe that incorporating the user and list ID embeddings into the final user and list representation is able to boost the performance, especially in Goodreads. This shows that our model captures some general characteristics of the user and list by the ID representation which benefits top-k list recommendation.
%	\item Remove Pre-training (Remove PT): The results show that our pre-training method cannot improve the performance significantly and even harms the performance in \textit{Goodreads} and \textit{Spotify}. Presumably this is because that word2vec model with the co-occurrence information of items is not a suitable initialization for top-k lists recommendation.

\smallskip
\noindent\textbf{+ Linear Projections:} Previous works  \cite{vaswani2017attention, kang2018self} apply three projection matrices to transform the input of self-attention layer into three matrices $Q$, $K$ and $V$.  Then, the scaled dot-product attention can be calculated as: $$ SelfAttention(Q, K, V) = softmax(\frac{QK^{T}}{\sqrt{d}})V$$ The projections make the model more flexible. However, we find that the projections make the network training more difficult and degrades the performance.%, especially in Zhihu with a drop of 6.2\% in P@10.
%\end{itemize}
%\end{description}

In summary, the vanilla attention network, the self-attention network with residual connections and the user and list ID representation are quite important for the architecture. Inspired by previous work, we also try to use position information \cite{liu2014recommending} and linear projections \cite{vaswani2017attention}. However, positional information of items has a limited impact while the length of a list is short (e.g., 32). Besides, self-attention networks should be carefully designed, where introducing linear projections increases a model's expressive capability but also makes the training more difficult.

%\subsection{Impact of Hyper-parameters}
\begin{figure}
  \centering
  \setlength{\abovecaptionskip}{0.1cm}
  \setlength{\belowcaptionskip}{-0.5cm}
  \subfigure[A Spotify playlist with high consistency among music tracks.]{
    \label{High internal similarities} %% label for first subfigure
    \includegraphics[width=1.5in]{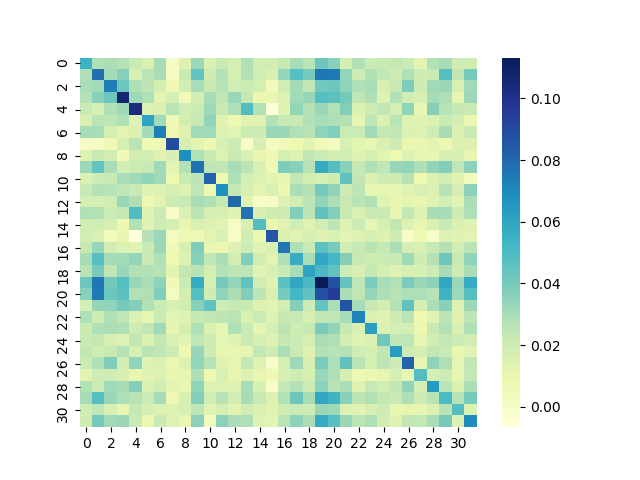}}
  %\hspace{1in}
  \subfigure[A Spotify playlist with low consistency among music tracks.]{
    \label{Low internal similarities} %% label for second subfigure
    \includegraphics[width=1.5in]{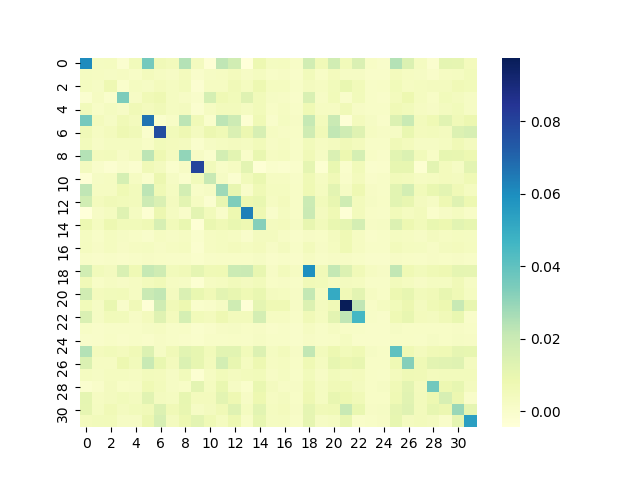}}
   \caption{Heatmap of self-attention scores of two lists.}
  \label{Heatmap} %% label for entire figure
\end{figure}

\subsection{RQ6: Impact of Hyper-parameters}
\label{RQ6: Impact of Hyper-parameters}
\begin{figure*}
  \centering
  \setlength{\abovecaptionskip}{0.1cm}
  \setlength{\belowcaptionskip}{-0.6cm}
  \subfigure[Latent dimensionality d]{
    \label{Latent dimensionality d} %% label for first subfigure
    \includegraphics[width=1.5in]{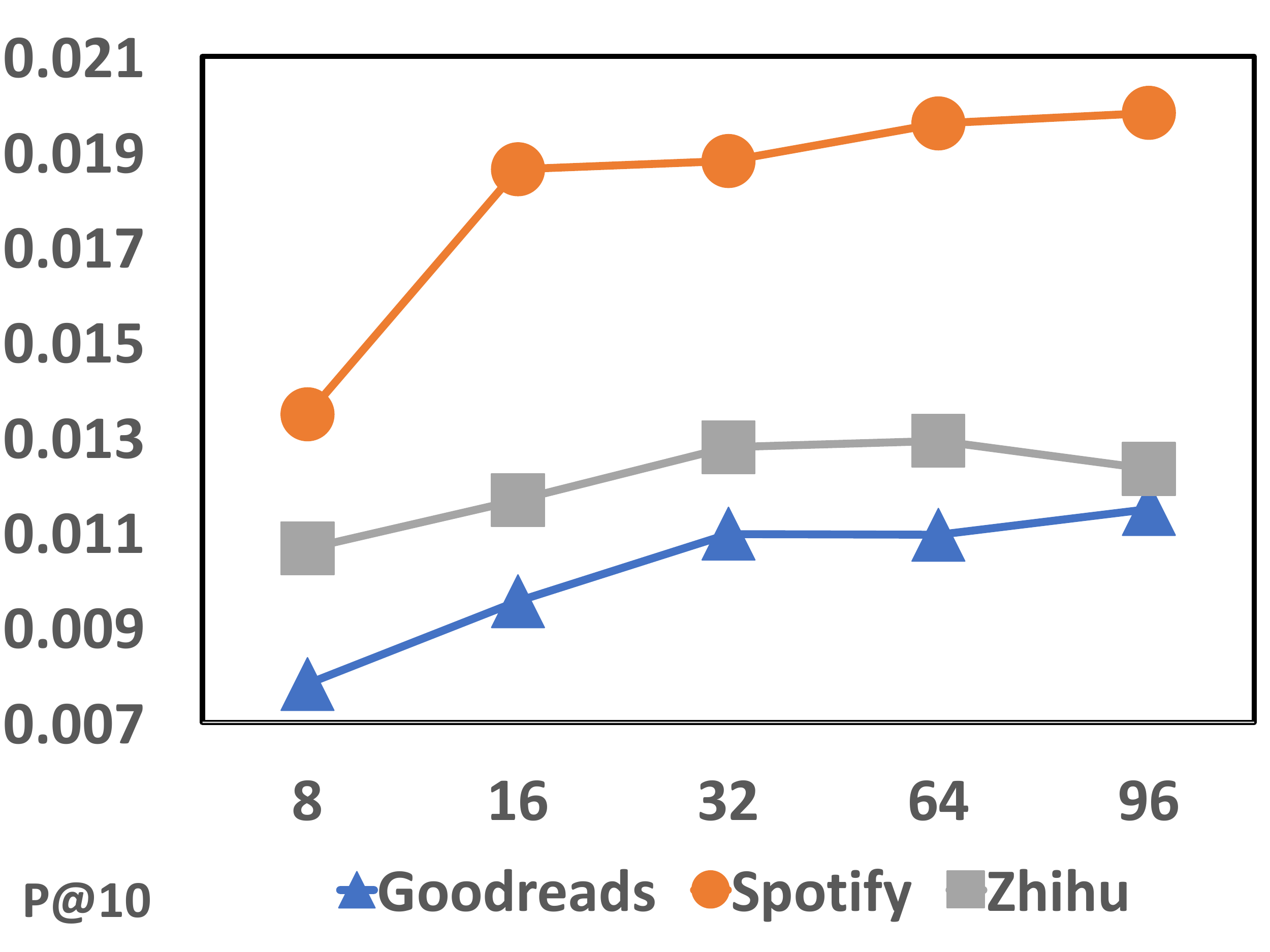}}
  %\hspace{1in}
  \subfigure[Number of items M]{
    \label{Number of items M} %% label for second subfigure
    \includegraphics[width=1.5in]{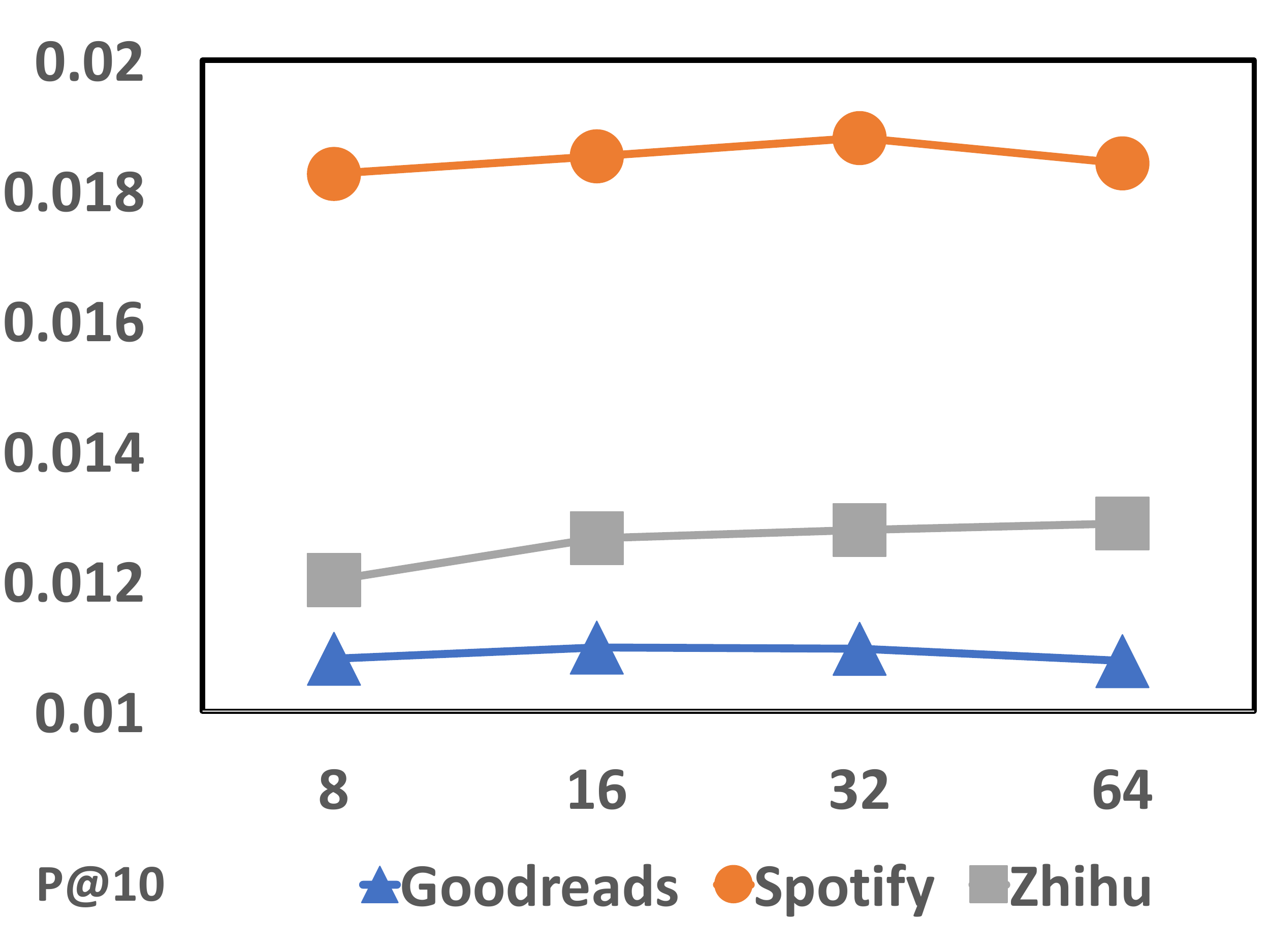}}
  \subfigure[Number of negatives $\rho$]{
    \label{Number of negatives} %% label for second subfigure
    \includegraphics[width=1.5in]{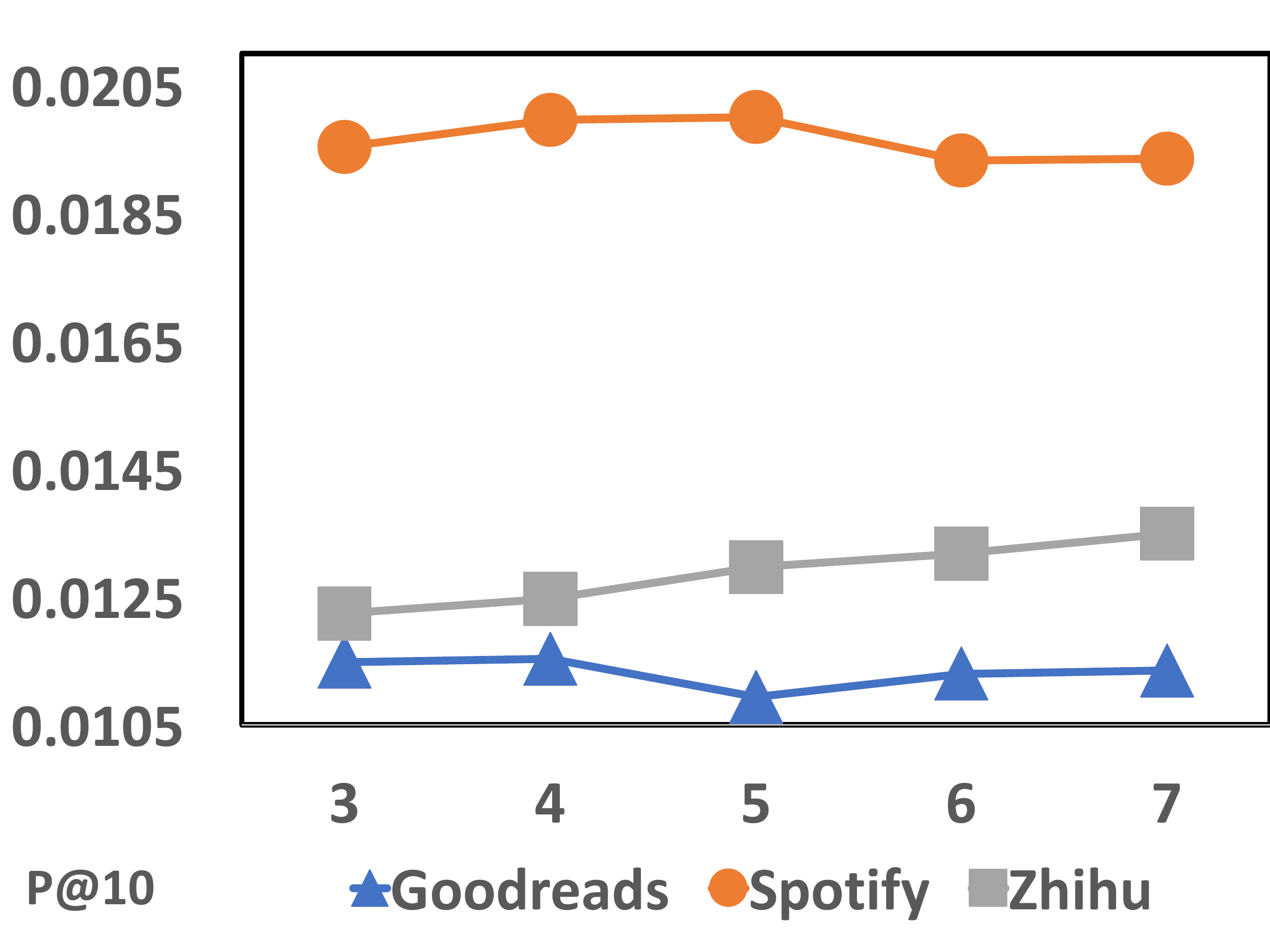}}
  \subfigure[Learning rate in Adam]{
    \label{Learning rate in Adam} %% label for second subfigure
    \includegraphics[width=1.5in]{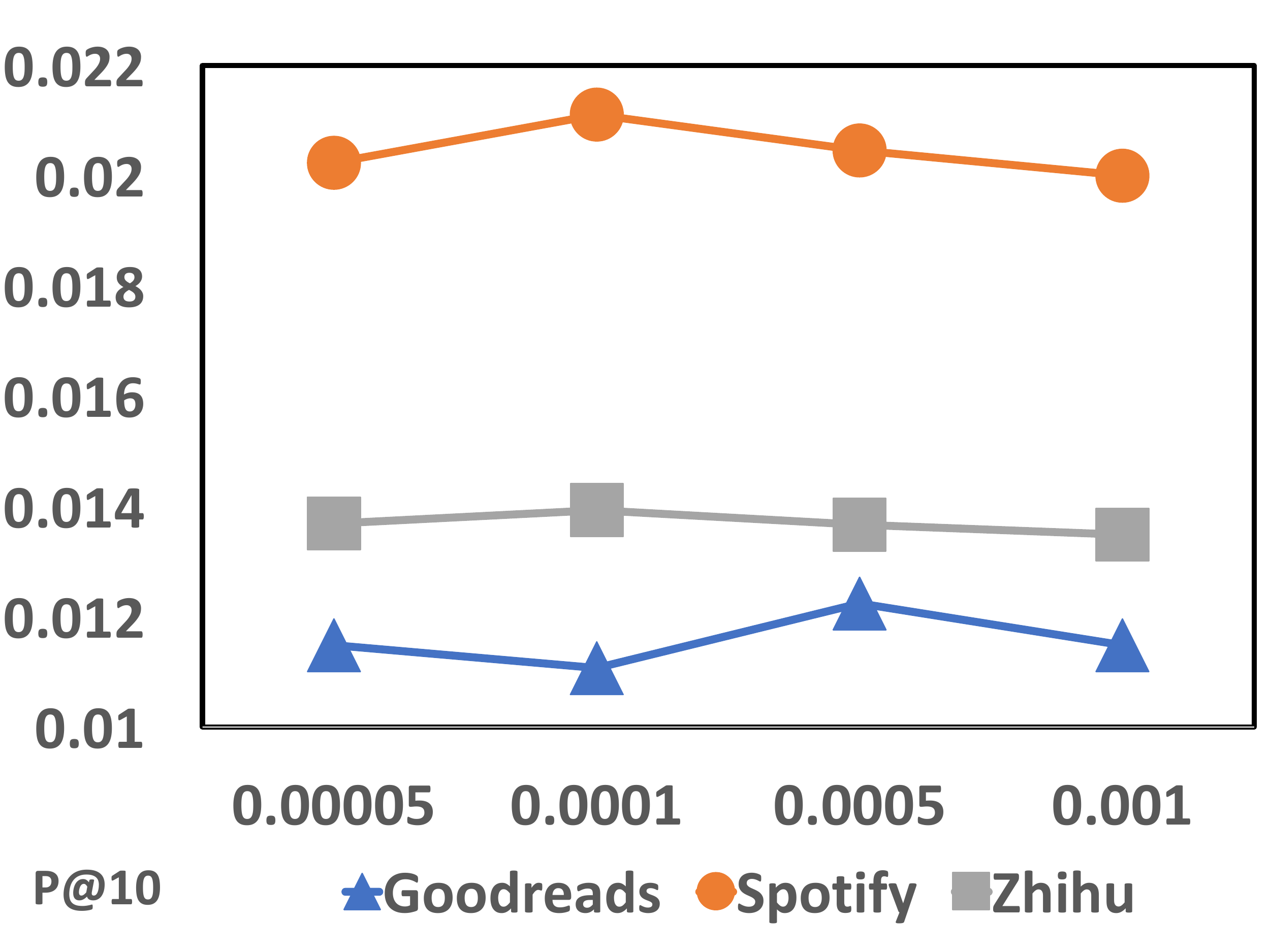}}

   \caption{Impact of the hyper-parameters of AttList.}
  \label{Impact of the hyper-parameters of AttList} %% label for entire figure
\end{figure*}

Due to the limited space, we focus here on four representative hyper-parameters of AttList to discuss their impact on performance. In Figure \ref{Latent dimensionality d}, we observe that AttList benefits from larger numbers of the latent dimensionality $d$. The best results are obtained with $d=96$ in Goodreads and Zhihu. In Figure \ref{Number of items M}, we surprisingly observe that a larger number ($M>32$) of items harms the performance in Spotify and Zhihu. Similarly, we find that a larger number ($N>15$) of lists harms the performance, which is omitted here for simplicity. Figure \ref{Number of negatives} shows that too many negative samples ($\rho>5$) harms the performance in Goodreads and Spotify. Figure \ref{Learning rate in Adam} shows that the learning rate should be tuned between 0.0001 and 0.0005 to obtain the best performance.

\subsection{RQ7: Visualizing Self-Attention Scores}
In this section, we visualize the self-attention scores and study if they can reflect consistency of neighboring items and lists as we claimed. As discussed in Section \ref{Self-attention networks section}, the self-attention score matrix in item-level aggregation layer is $F_{l} \in \mathbb{R}^{M \times M}$, where each element is the similarity between two items in the list. We pick two typical lists from Spotify and plot the heatmap of self-attention scores of songs within the two lists as shown in Figure \ref{Heatmap}, where the darker the color is, the higher the self-attention score is. Two observations can be obtained from Figure \ref{Heatmap}. First, the diagonal elements in the two matrices have higher self-attention scores. This is understandable that each item is similar to itself, which reaffirms that the self-attention scores are actually the similarities between one item and its neighboring items. Second, as a whole, the heatmap in Figure \ref{High internal similarities} is darker than the heatmap in Figure \ref{Low internal similarities}. This illustrates that the list in \ref{High internal similarities} has a higher internal consistency than the list in \ref{Low internal similarities}. A possible reason for that is the first list curates songs more homogeneous (e.g., similar genre) than the second list. This case shows that our self-attention network is able to capture consistency among items within a list. %A possible reason for that is the list in \ref{High internal similarities} are more consistent in 

\begin{figure}
  \centering
  \setlength{\abovecaptionskip}{0.1cm}
  \setlength{\belowcaptionskip}{-0.6cm}
  \subfigure[Self-attention scores for the high consistency Spotify playlist.]{
    \label{song high internal similarities} %% label for first subfigure
    \includegraphics[width=1.5in]{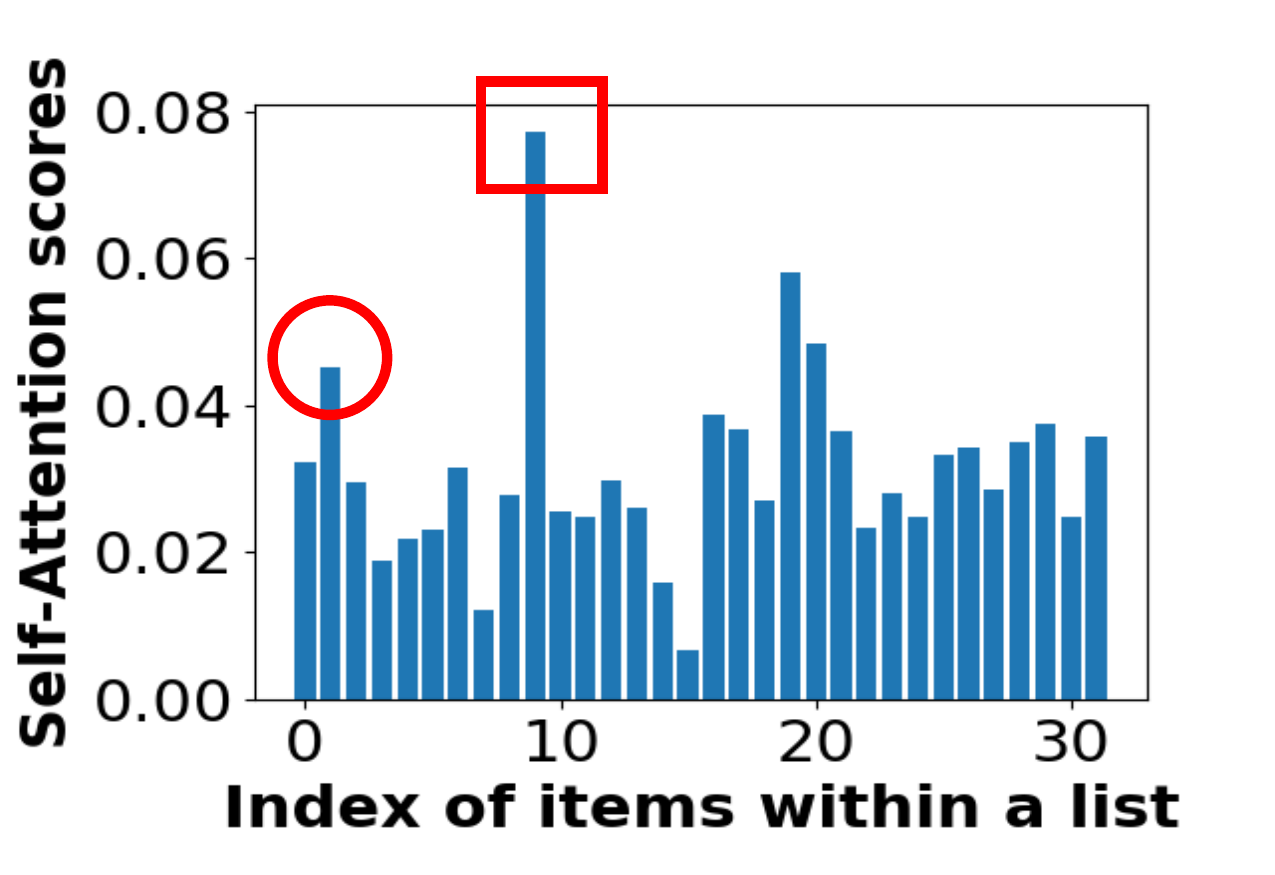}}
  %\hspace{1in}
  \subfigure[Self-attention scores for the low consistency Spotify playlist.]{
    \label{song low internal similarities} %% label for second subfigure
    \includegraphics[width=1.5in]{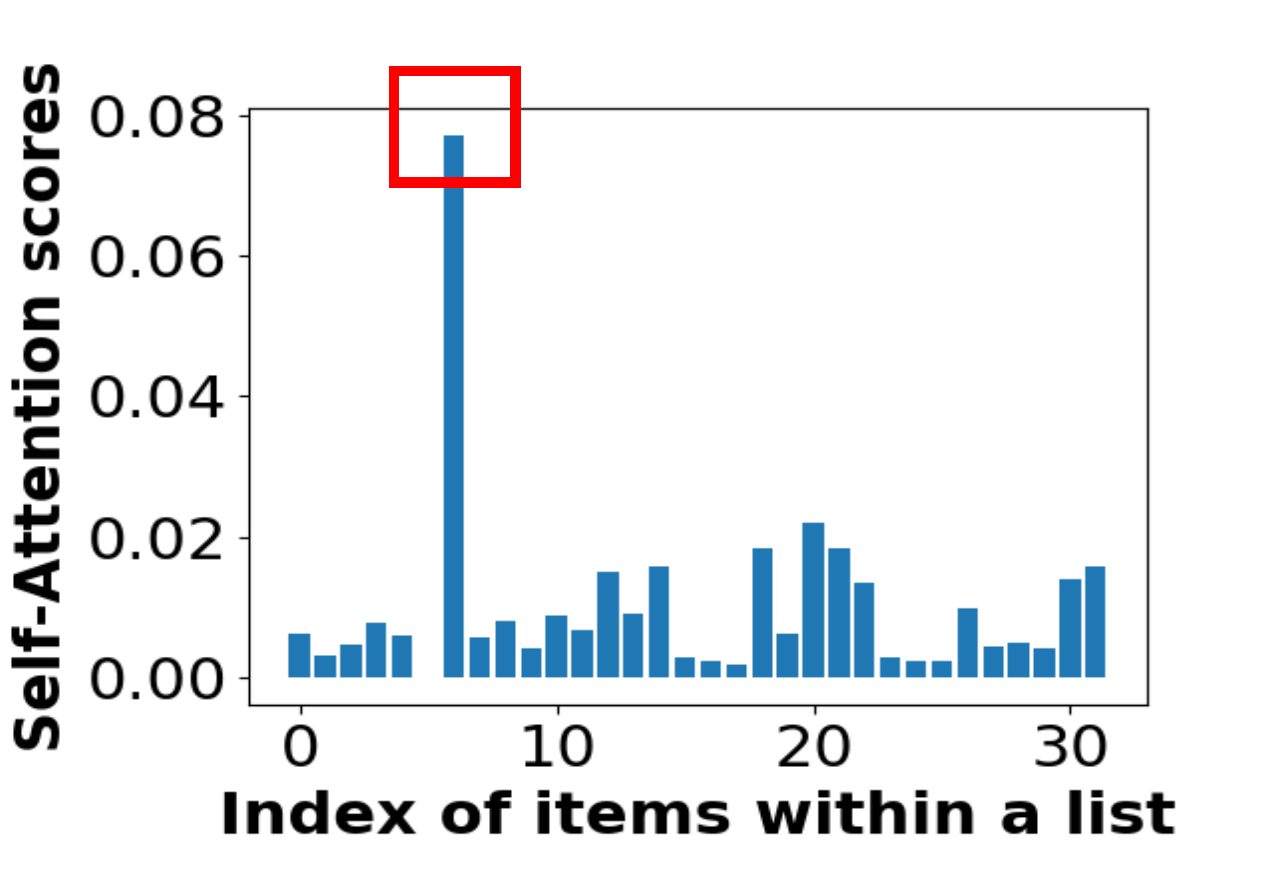}}
   \caption{The self-attention scores of Taylor Swift's \textit{Safe and Sound} on the two lists also featured in Figure \ref{Heatmap}. The square marks \textit{Safe and Sound} and the circle marks \textit{Shake It Off}.}
  \label{self-attention bar} %% label for entire figure
\end{figure}

To further investigate the self-attention scores, we present here a case study for a well-known song \textit{Safe and Sound} by \textit{Taylor Swift}, which appears in the two lists in Figure \ref{self-attention bar}. We show the self-attention scores of it in the high internal consistence list in Figure \ref{song high internal similarities} and the scores in the low internal consistence list in Figure \ref{song low internal similarities}. We observe that \textit{Safe and Sound} has higher self-attention scores with other songs in \ref{song high internal similarities} than in \ref{song low internal similarities}, which is consistent with the observation from Figure \ref{Heatmap}. Taking a song \textit{Shake It Off} in Figure \ref{song high internal similarities} as an example, this song has a relatively high self-attention score (0.045, the maximum score is 0.072 in the list) to \textit{Safe and Sound}. Interestingly, the two songs are both performed by \textit{Taylor Swift}. This illustrates that the semantic relationship among songs can be captured by self-attention scores.

% Table generated by Excel2LaTeX from sheet 'goodreads_latex'
\section{Conclusions and Future Work}
In this paper, we tackle the problem of recommending user generated item lists through a new hierarchical self-attentive recommendation model. Unlike traditional collaborative filtering based algorithms that are optimized for user-list interactions, our proposed model leverages the hierarchical structure of items, lists, and users to capture the containment relationship between lists and items, revealing additional insights into user preferences. With this hierarchical structure among items, lists, and users, our model first aggregates items to characterize the lists they belong to, and then aggregates these lists to estimate user preferences. A key aspect of the proposed approach is a novel self-attention network to refine the item representations and list representations by considering consistency of neighboring items and lists. Experiments over three real-world domains -- Goodreads, Spotify, and Zhihu -- demonstrate the effectiveness of AttList versus state-of-the-art item-based and list-based recommenders. Furthermore, these datasets and code will be released to the research community for further exploration.%To evaluate our model, we construct three datasets by crawling and cleaning data from three real-world applications: Goodreads, Spotify and Zhihu, which will be released to research community for further exploration. Experimental results demonstrate that our model is significantly superior to other baselines.

In our continuing work, we plan to improve AttList in the following two directions: 1) We plan to use list titles (e.g., ``my favorite adventure books") for improving our model, which also contain rich signals of user preference. 2) The social community around a list also has a considerable impact on the behavior of interactions between users and lists. For example, users are more likely to interact with lists that their friends have interacted with. Thus, the social influence can be modeled to improve the performance.
\label{Conclusion and Future Work}

\begin{acks}
This work is supported in part by NSF (\#IIS-$1841138$). %The views, opinions, and/or findings expressed are those of the author(s) and should not be interpreted as representing the official views or policies of the Department of Defense or the U.S. Government.
\end{acks}

\bibliographystyle{ACM-Reference-Format}
\bibliography{sample-bibliography}

\end{document}